\documentclass[aps,prb,preprint]{revtex4}
\usepackage{amsmath,amsfonts,graphicx}

\begin{document}

\title{Spectral properties of exciton polaritons in one-dimensional resonant photonic
crystals}

\author{M. V. Erementchouk}
\author{L. I. Deych}
\author{A. A. Lisyansky}

\affiliation{Physics Department, Queens College, City University
of New York, Flushing, New York 11367, USA}

\begin{abstract}
The dispersion properties  of exciton polaritons in
multiple-quantum-well based resonant photonic crystals are studied.
In the case of structures with an elementary cell possessing a
mirror symmetry with respect to its center, a powerful analytical
method for deriving and analyzing dispersion laws of the respective
normal modes is developed.  The method is used to analyze band
structure and dispersion properties of several types of resonant
photonic crystals, which would not submit to analytical treatment by
other approaches. These systems include multiple quantum well
structures with an arbitrary periodic modulation of the dielectric
function and structures with a complex elementary cell. Special
attention was paid to determining conditions for superradiance
(Bragg resonance) in these structures, and to the properties of the
polariton stop band in the case when this condition is fulfilled
(Bragg structures). The dependence of the band structure on the
angle of propagation, the polarization of the wave, and the effects
due to exciton homogeneous and inhomogeneous broadenings are
considered, as well as dispersion properties of excitations in
near-Bragg structures.
\end{abstract}

\maketitle

\section{Introduction}

Optical properties of artificial structures with periodically
modulated dielectric constant has been attracting a great deal of
interest since pioneering papers
Refs.~\onlinecite{YABLONOVITCH:1987} and \onlinecite{JOHN:1987},
where such systems have been first discussed. Periodic modulation
of the dielectric function significantly modifies spectral
properties of electromagnetic waves. Instead of a simple
continuous spectrum with a linear dispersion law, the
electromagnetic spectrum in such structures  is characterized by
the presence of allowed and forbidden bands similar to electronic
band structure of crystals. For this reason, the new class of
optical materials was dubbed photonic
crystals.\cite{Joannopoulos_PC_MFL:1995} Changing the spatial
distribution of the dielectric constant one can effectively
control such fundamental properties as the group velocity of
light, the rate of spontaneous emission, etc. This fact has
important repercussion for both fundamental physics and for
applications, where it opens up possibilities for new concepts of
optical and optoelectronic devices.

Modulation of the dielectric function, however, is not the only
way to affect light propagation and its interaction with matter.
It was shown in Ref.~\onlinecite{DICKE_PR:1954} that closely
packed dipole active atoms can become coherently coupled by light,
and this coupling significantly changes their emission properties
resulting in the so called ``super-radiance" effect. In the
original Dicke's model\cite{DICKE_PR:1954} the distance between
the atoms was assumed to be much smaller than the wavelength of
their emission, but similar effect can also arise if dipole active
elements form a one-dimensional periodic lattice with the period
coinciding with the emission half-wavelength (Bragg resonance).
Such an arrangement is possible with a so called optical lattice
of cold atoms\cite{DEUTSCH:1995} and with their semiconductor
analogs -- Bragg multiple quantum well (MQW)
structures.\cite{Keldysh,IvchenkoMQW,IvchenkoSpectrum,KhitrovaMQWprl}
The latter are semiconductor heterostructures, in which very thin
layers of a narrower band gap semiconductor (quantum well) are
separated from each other by much thicker layers of a wider band
gap semiconductor (barrier) in such a way that the period of the
structure coincides with the half-wavelength of light emitted by
excitons confined in a quantum well (QW).  In these systems,
excitons play the role of the dipole active excitations, which
become radiatively coupled and demonstrate the superradiance
effect. If, however, the size (the number of periods) in an
optical lattice increases, the properties of the system
change\cite{DLPRBSpectrum,IKAWA:2002}. In particular, the dark
modes form two branches of collective excitations, in which light
is coupled with the material resonances, and which can be called
optical lattice polaritons. At the same time, the super-radiance
mode develops into a stop band, which is a spectral gap between
the two polariton branches, characterized by almost complete
reflection of a normally incident radiation. The distinction
between Bragg structures and other arrangements, in which the
period of the structure is not in the resonance with radiation
emitted by the active elements, can, in this case, be described in
terms of formation of a significantly (by orders of magnitude)
enhanced stop band.

There is, however, an important difference between optical lattices
of atoms and semiconductor heterostructures. In the latter case, the
interaction of light with active elements (excitons) is accompanied
by multiple reflection of light from interfaces between wells and
barrier layers caused by difference in their refractive indexes.
Thus, in structures like MQWs optical lattice effects coexist with
photonic crystal-like effects, which results in a number of new and
interesting optical properties and opportunities for applications.
Such structures, called resonant photonic crystals (RPC), have
recently started
attracting significant attention.\cite{KUZMIAK:1997,DEYCH:1998,NOJIMA:2000,%
Raikh_braggaritons,CHRIST:2003,HUANG:2003,TOADER:2004,PILOZZI:2004}
A general characteristic of this type of structures is the
co-existence of dipole active material excitations (excitons,
phonons, plasmons) described by optical susceptibilities of the
resonant type and the periodic modulation of the background
dielectric constant. Two most fundamental problems in the focus of
current research in this area are concerned with the effects of
the interplay between resonances and spatial non-uniformity of the
dielectric function on the band structure of these systems and
their optical spectra. While these two questions are
interconnected, they have to be recognized as two separate
problems, one dealing with normal modes of closed (or infinite)
systems and their dispersion laws and the other with the
interaction of an internal radiation with a finite size structure.

The main interest of studies in this field is a spectral region in
the vicinity of the resonant frequency, where the interplay
effects play the most important role. A theoretical description of
RPC's in this spectral region is a challenging problem and its
complete solution has not yet been obtained even for a simplest
case of one-dimensional structures, such as MQWs. Of course, it is
always possible to carry out numerical calculations of the optical
spectra and the dispersion laws, which are particularly easy in
one-dimensional case. However purely numerical approach does not
provide a real understanding of physics of these structures.
Therefore, it is very important to be able to carry out analytical
analysis, which would not only provide a better understanding of
physical processes taking place in these structures, but would
also be instrumental in designing structures with pre-determined
properties, which is crucial for their possible applications.

Different aspects of this problem have been, of course, considered
in a number of previous publications. In particular, in
Ref.~\onlinecite{IvchenkoContrast} the necessity of a modification
of the Bragg resonance (super-radiance) condition for
one-dimensional resonant photonic crystals compared to the case of
optical lattice structures was shown. Later, this result was
confirmed\cite{JointComplex} and an exact Bragg condition in such
structures was found for the particular case of normal propagation
of the electromagnetic wave. It was also noted in
Ref.~\onlinecite{JointComplex} that, when the resonance condition
is met, the spectral gap between the polariton branches becomes
wider than in the case of a passive photonic crystal characterized
by the same modulation of the dielectric function or of an optical
lattice with the same strength of radiative coupling between the
active elements. In the case of structures of higher (2 or 3)
dimensions, no analytical results are available yet, but one can
note recent numerical calculations presented in
Refs.~\onlinecite{TOADER:2004} and \onlinecite{IVCHENKO:2005},
where it was also found that the resonances may enhance polariton
related band gaps compared to a purely passive photonic crystal
with the same spatial distribution of the dielectric constant.
It is important to emphasize that this enhancement of the band gap
in RPC structures is not a trivial effect. In order to illustrate
this point we can mention another, in a certain sense, opposite
effect, which also results from the interplay between the
resonances and periodic inhomogeneities of the dielectric
function. It was shown in
Ref.~\onlinecite{Optical_Properties_MQWPC} that there always
exists a frequency where the reflection coefficient of the
incident light vanishes due to the destructive interference
between the two channels of interaction between light and the
structure.


These examples show the richness of the optical properties of even
one-dimensional resonant  photonic crystals. In order to achieve a
complete understanding of relationships between optical and
geometrical characteristics of these structures, one needs a
general analytical method, which could be applied to a variety of
RPCs independently of a particular spatial dependence of their
dielectric function. In our previous paper,
Ref.~\onlinecite{Optical_Properties_MQWPC}, such a method was
developed for studying \emph{reflection, transmission, and
absorption spectra} of one-dimensional RPC structures. The present
paper presents a general analytical approach to calculating the
\emph{band structure} of these materials. This method allows one
to analyze dispersion laws of one-dimensional RPCs with an
arbitrary form of the periodic modulation of the dielectric
function, propagation angle of the electromagnetic wave and its
polarization state. This method
naturally incorporates resonant excitations of the medium into the
theory. For concreteness, we assume that the resonances are related
to exciton states in QWs, but this approach can be easily applied to
any other type of the resonance excitations. In
Sections~\ref{sec:derivation_transfer_matrix} and, particulary, in
subsection~\ref{sec:structure_transfer_matrices} we show that by
choosing an appropriate basis for representing transfer matrices one
can significantly simplify the analysis of the band structure and
dispersion laws of electromagnetic excitations in the structures
under consideration. In Sections IV and V we illustrate the power of
the approach by applying it to a number of examples, some of which
deal with already well studied situations (passive one-dimensional
photonic crystals, optical lattices, MQWs with a piece-wise spatial
dependence of the dielectric function), while the others demonstrate
the possibilities of our approach in obtaining results that could
not have been obtained by other methods.

\section{The $(a,f)$-representation of the transfer matrix}
\label{sec:derivation_transfer_matrix}
\subsection{Transfer matrix approach in resonant photonic crystals}
\begin{figure}
  \includegraphics[width=3 in]{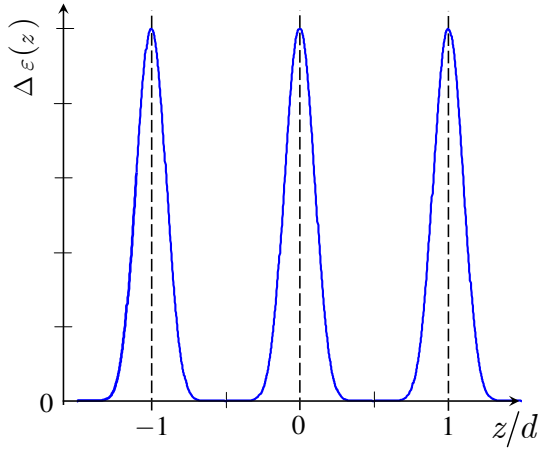}\\
  \caption{An example of the modulation of the dielectric
function,
  $\epsilon(z) = \bar{\epsilon}+\Delta\epsilon(z)$
  (solid line). The dashed vertical lines
  show the positions of the centers of QWs.}\label{fig:general_modulation}
\end{figure}

A propagation of the electromagnetic wave in structures under
discussion is governed by the Maxwell equation
\begin{equation}\label{eq:Maxwell_equations}
  \nabla \times \nabla \times \mathbf{E} = \frac{\omega^2}{c^2}[
  \epsilon_\infty(z) \mathbf{E} + 4\pi \mathbf{P}_{exc}],
\end{equation}
where $z$ axis is chosen along the growth direction and
$\mathbf{P}_{exc}$ is the excitonic contribution to the
polarization, which can be presented in the following form:
\begin{equation}\label{eq:exciton_polarization}
  \mathbf{P}_{exc} = -\chi(\omega)\sum_m\Phi_m(z)
  \int dz\, \Phi_m(z')\mathbf{E}_{\perp}(z'),
\end{equation}
where $\Phi_m(z) = \Phi(z-z_m)$ is the envelope wave function of
an exciton localized in the $m$-th QW. The summation in
Eq.~(\ref{eq:exciton_polarization}) is performed over all QWs and
$z_m$ are the positions of their centers. We assume that the
distance between the consecutive wells, $d=z_{m+1} - z_m$,
coincides with the period of the spatial modulation of the
dielectric function, $\epsilon(z+d) = \epsilon(z)$. Also, we
assume that the profile of the dielectric function is symmetrical
with respect to the position of the center of the QW,
$\epsilon(z_m + z) = \epsilon(z_m - z)$ (see
Fig.~\ref{fig:general_modulation}). We restrict ourselves to the
consideration of $1s$ states of heavy-hole excitons and neglect
their in-plane dispersion. The dipole moment of these excitons
lies in the plane of the well, therefore, they can only interact
with the in-plane component of the electric field,
$\mathbf{E}_{\perp}$. This fact is reflected in
Eq.~(\ref{eq:exciton_polarization}), where only this component of
the field is included.

The frequency dependence of the excitonic response is given by the
function, $\chi(\omega)$, defined as
\begin{equation}\label{eq:susceptibility_ideal}
  \chi(\omega) = \frac{\alpha}{\omega_0 - \omega - i\gamma},
\end{equation}
where $\omega_0$ is the exciton resonance frequency, $\gamma$ is the
non-radiative decay rate of the exciton, and $\alpha$ is the
microscopic exciton-light coupling parameter, proportional to the
dipole moment of the electron -- heavy hole transition.

Eq.~(\ref{eq:Maxwell_equations}) can be analyzed by presenting the
direction of the electric field as a vector sum of two mutually
perpendicular polarizations. These, so called $s$- and
$p$-polarizations, define two possible eigen directions, and
accordingly, the respective electric fields satisfy two
independent equations, which can be considered separately. In the
structures modulated in the $z$-direction, normal modes of the
system are characterized by the wave vector perpendicular to the
modulation (growth) direction, which is a conserving quantity. The
direction of this vector can be also used to define the eigen
polarizations of the waves.

The absence of an overlap of the exciton wave functions localized
in different QWs makes it possible to use the transfer matrix
technique for solving Maxwell equations,
Eq.~(\ref{eq:Maxwell_equations}). The general idea of this
technique is to characterize the electric field of the wave with a
certain polarization by a two-component column vector,
$|c(z)\rangle$ whose change along the structure is governed by a
$2\times 2$ matrix. The main property of this matrix can be
described as following. If $|c(z_1)\rangle$, $|c(z_2)\rangle$, and
$|c(z_3)\rangle$ characterize the field at points $z_1$, $z_2$,
and $z_3$, respectively, and if $\widehat{T}(z_i,z_j)$ satisfies
the equation $|c(z_i)\rangle=\widehat{T}(z_i,z_j)|c(z_j)\rangle$,
then $\widehat{T}(z_1,z_3)=
\widehat{T}(z_1,z_2)\widehat{T}(z_2,z_3)$. The relations between
the values of the field and the vector $|c(z)\rangle$ as well as
the form of the matrix $\widehat{T}(z_i,z_j)$ depend on the choice
of the particular representation of the field. Several possible
representations will be discussed in the subsequent sections of
the paper for both polarizations.


\subsubsection{$S$-polarization}

In the $s$-polarized wave the electric field $\mathbf{E}$ is
perpendicular to the plane formed by the direction of $z$ axis and
in-plane wave vector, $\mathbf{k}$. Accordingly, it can be presented
in the form
\begin{equation}\label{eq:field_representation_s_polarization}
  \mathbf{E}(z, \boldsymbol{\rho}) = \hat{\mathbf{e}}_s
  E(z)e^{i \mathbf{k}\boldsymbol{\rho}},
\end{equation}
where $\boldsymbol{\rho}$ is the coordinate in the $(x,y)$-plane
and $\hat{\mathbf{e}}_s = \hat{\mathbf{e}}_\mathbf{k}\times
\hat{\mathbf{e}}_z$ is a unit polarization vector, defined in
terms of unit vectors in $z$- and $\mathbf{k}$-directions. For
$E(z)$ we obtain an integro-differential equation
\begin{equation}\label{eq:Maxwell_smooth_modulation_s_pol}
  \frac{d^2E(z)}{dz^2}+\kappa_s^2(z) =
-\chi(\omega)\frac{4\pi\omega^2}{c^2}
  \sum_m\Phi_m(z) \int dz\, \Phi_m(z')E(z'),
\end{equation}
where $\kappa_s^2(z) = \omega^2 \epsilon(z)/c^2 - k^2$.

We first consider the transfer matrix corresponding to the
propagation of the field across one elementary cell of the
structure, i.e. from point $z_-+0$ at the left boundary of the
cell, inside it, to the point $z_++0$ just outside of the right
boundary of the cell. The complete transfer matrix connecting the
field at the right boundary of the entire system with the field at
its left boundary can then be obtained as a product of the
transfer matrices for each cell. Considering only one cell, we
can, without any loss of generality, choose the origin of our
coordinate system coinciding with a position of the QW in the cell
under consideration. When $z$ coordinate is confined to a single
cell, the summation over QWs and the index enumerating them in
Eq.~(\ref{eq:Maxwell_smooth_modulation_s_pol}) can be dropped.

The term with the exciton polarization at the r.h.s. of
Eq.~(\ref{eq:Maxwell_smooth_modulation_s_pol}) can be considered
as an inhomogeneity in a second order differential equation
\begin{equation}\label{eq:inhomogeneous_equation}
 \frac{d^2E(z)}{dz^2}+ \kappa_s^2(z)E = \mathcal{F}(z),
\end{equation}
whose general solution can be written in the
form\cite{Morse_Feshbach}
\begin{equation}\label{eq:general_solution}
  E(z) =  c_1 h_1(z) + c_2 h_2(z) + (G \star \mathcal{F})(z) ,
\end{equation}
where
\begin{equation}\label{eq:PC_Green_function_s_pol}
  (G \star \mathcal{F})(z) = \int_{z_-}^z dz'
\mathcal{F}(z')\frac{h_1(z')h_2(z)-h_1(z)h_2(z')}{W(h_1,h_2;z')}.
\end{equation}
Here we have introduced $h_{1,2}(z)$, a pair of linearly independent
solutions of the homogeneous equation
\begin{equation}\label{eq:pure_PC_equation}
  \frac{d^2E(z)}{dz^2}+ \kappa_s^2(z)E = 0,
\end{equation}
and $W(h_1,h_2;z) = h_1h_2'-h_1'h_2$ is the Wronskian of these
solutions. For the case under consideration the Wronskian does not
depend on $z$ and will be denoted by $W_h$ in what follows.

Alternatively, a solution of
Eq.~(\ref{eq:Maxwell_smooth_modulation_s_pol}) can be presented as
\begin{eqnarray}\label{eq:altern_represnt}
E(z) &= & c_1(z) h_1(z) + c_2(z) h_2(z),\\
E^\prime(z) &= & c_1(z) h_1^\prime(z) + c_2(z)
h_2^\prime(z).\nonumber
\end{eqnarray}
In the regions outside of the QW, where $\Phi(z)\equiv 0$,
functions $c(z)$ remain constants, making representations given by
Eq.~(\ref{eq:general_solution}) and Eq.~(\ref{eq:altern_represnt})
equivalent, but $c_i(z)$ change when $z$ traverses from one
boundary of the QW to the other. Our goal now is to use
Eq.~(\ref{eq:general_solution}) along with
Eq.~(\ref{eq:PC_Green_function_s_pol}) in order to relate the
values of $c_i(z)$ at the right boundary of the QW to their values
at the left boundary. Using these equations we can obtain the
expression for the electric field for points to the right from the
QW as
\begin{equation}\label{eq:general_solution_explicit}
\begin{split}
 E(z) = &h_1\left[c_1 +
\tilde\chi\frac{4\pi\omega^2\varphi_2}{c^2}
       \left(c_1 \varphi_1 + c_2 \varphi_2 \right)\right]\\
 +&h_2\left[c_2 - \tilde\chi\frac{4\pi\omega^2\varphi_1}{c^2}
\left(
        c_1 \varphi_1 + c_2 \varphi_2)\right)\right],
\end{split}
\end{equation}
where $\varphi_{1,2}$ are ``projections" of the solutions
$h_{1,2}$ onto the exciton states
\begin{equation}\label{eq:projection_of_PC_modes}
  \varphi_{1,2} = \frac{1}{W_h}\int_{QW}dz' \Phi(z')h_{1,2}(z').
\end{equation}
and the modified excitonic response function $\tilde\chi$ can be
presented as
\begin{equation}\label{eq:modified_susceptibility}
  \tilde\chi = \frac{\chi}{1- \Delta\omega\chi/\alpha},
\end{equation}
where
\begin{equation}\label{eq:radiative_shift}
  \Delta\omega = \alpha\int_{QW}dz\, \Phi(z)(G\star\Phi)(z)
\end{equation}
gives the radiative shift of the exciton frequency in the photonic
crystal. Eq.~(\ref{eq:radiative_shift}) is a generalization of the
well-known expression for the radiative shift in MQW structures with
a homogeneous dielectric
function.\cite{AndreaniPRB1992,AndreaniKavokinIB,IKAWA:2002}

Comparing Eq.~(\ref{eq:general_solution_explicit}) with
Eq.~(\ref{eq:altern_represnt}), and taking into account that the
coefficients $c_i$ do not change outside of the QW, we can find a
relation between the coefficients $c_i$ determined at the two
\emph{inner} boundaries of the elementary cell, $z=z_{-}+0$, and
$z=z_{+}-0$
\begin{equation}\label{eq:amplitudes_transfer}
  \begin{pmatrix}
    c_1 \\
    c_2
  \end{pmatrix}(z_{+}-0) = T_h \begin{pmatrix}
    c_1 \\
    c_2
  \end{pmatrix}(z_{-}+0),
\end{equation}
with matrix $T_h$ given as
\begin{equation}\label{eq:transfer_matrix_functions_basis}
  T_h = \boldsymbol{1} + \frac{4\pi\omega^2\tilde\chi}{c^2}
  \begin{pmatrix}
    \varphi_2\varphi_1 & \varphi_2^2 \\
    -\varphi_1^2 & -\varphi_2\varphi_1
  \end{pmatrix},
\end{equation}
where $\boldsymbol{1}$ is the unit matrix. It should be noted that
Eq.~(\ref{eq:transfer_matrix_functions_basis}) is valid for an
arbitrary form of the exciton envelope wave function and spatial
modulation of the dielectric function.

Eq.~(\ref{eq:transfer_matrix_functions_basis}) can be
significantly simplified if we use freedom in the choice of the
functions $h_i(z)$. In the case when $\epsilon(z)$ is invariant
with respect to the mirror reflection relatively to the center of
the QW, $h_{1,2}$ can be chosen to have a definite parity, with
one of them being even with respect to the center of the QW, and
the other being odd. With such a choice of these functions we can
always turn one  of $\varphi_{1,2}$ to zero.\footnote{It raises a
question is it possible to have both $\varphi_{1,2}$ equal to $0$.
Generally it can be proven that it is impossible below the
frequency of the first photonic band gap, so the latter is always
effected by the exciton-light interaction. This impossibility can
be proven also for all frequencies for considered here model of
symmetric QWs and spatial modulation of the dielectric function,
and for $\delta$-functional approximation for the envelope wave
function. These results make the question about the possibility
for the effective exciton-light interaction to be completely
inhibited rather academic.}
It should be understood, however, that $h_1$ and $h_2$ are not
actual normal modes of the photonic crystal. The latter must be
defined as solutions of the appropriate boundary problem, and do not
have to be even or odd functions with respect to the center of the
elementary cell. Moreover, due to the Bloch theorem the modes of a
photonic crystal can  have a definite parity only at specific
frequencies that are naturally identified with boundaries of the
forbidden gap in the spectrum. However, all normal modes of the
photonic crystal can be represented as superpositions of the even
and odd solutions $h_{1,2}$. A formal discussion of the relation
between the functions $h_1$ and $h_2$ and the normal modes can be
found in Ref.~\onlinecite{Vainberg_delays} where the similar
approach has been used for the analysis of spectral properties of a
Schr\"{o}dinger equation with a periodic potential.

Obviously, fixing the parity of the solutions does not determine
functions $h_1$ and $h_2$ uniquely since a multiplication by a
constant determined by initial conditions does not change their
symmetry. It can be shown, however, that results obtained for
observable quantities, such as band structure, do not depend on this
ambiguity. We will present our general results in a form independent
on the choice of initial conditions. However, for discussions of
particular examples it is more convenient to fix them in the form
\begin{equation}\label{eq:initial_conditions_for_fs}
\begin{split}
 h_1(0) = 1, \qquad h_1'(0) = 0 , \\
 h_2(0) = 0, \qquad h_2'(0) = 1,
\end{split}
\end{equation}
for which $W_h=1$.

Let us assume that $h_1$ is the even solution, then $\varphi_2
\equiv 0$ and the expression for the matrix $T_h$ simplifies to
\begin{equation}\label{eq:transfer_matrix_functions_basis_simple}
  T_h = \boldsymbol{1} + S_s q_s
   \begin{pmatrix}
    0 & 0 \\
    1 & 0
  \end{pmatrix},
\end{equation}
where $q_s = \kappa_s(z_+)$ is the value of $\kappa_s(z)$ at the
boundary of the elementary cell and
\begin{equation}\label{eq:S_def_s_polarization}
  S_s = - \tilde{\chi}(\omega)\frac{2\pi\omega^2\varphi_1^2}{q_s c^2 W_h}.
\end{equation}
Substitution of $\tilde{\chi}$ yields
\begin{equation}\label{eq:Lorentzian_S_s_polarization}
  S_s = \frac{\Gamma_s}{\omega - \omega_0 - \Delta\omega +  i\gamma},
\end{equation}
where $\Gamma_s$ is the radiative decay rate,
\begin{equation}\label{eq:Gamma_s_def}
  \Gamma_s = \frac{2\pi\alpha \omega^2\varphi_1^2}{q_s c^2}.
\end{equation}
The function $S_s(\omega)$, which we shall refer to as exciton
susceptibility, describes the contribution of the exciton-light
interaction to the optical properties of the structures under
consideration. For instance, the resonant absorption of light
occurs at the frequency $\tilde{\omega}_0=\omega_0 + \Delta\omega$
determined by the pole of $S_s(\omega)$. We will treat
$\tilde{\omega}_0$ as an experimentally accessible exciton
frequency, which along with the radiative decay rate $\Gamma_s$
can be measured in optical experiments with a single QW. In what
follows we will drop the tilde from $\omega_0$ and assume that the
radiative shift is included in this parameter.

Coefficients $c_i$ in Eq.~(\ref{eq:amplitudes_transfer}) are the
two components of the vector $|c\rangle$ characterizing the field
in the transfer matrix formalism  while matrix $T_h$ given by
Eq.~(\ref{eq:transfer_matrix_functions_basis}) is the transfer
matrix itself. The functions $h_i(z)$, which we used to derive
Eqs.~(\ref{eq:amplitudes_transfer}) and
(\ref{eq:transfer_matrix_functions_basis_simple}) form a basis for
this particular representation of the transfer matrix. Using this
basis we are able to obtain a simple and convenient expression for
the transfer matrix describing the evolution of the electric field
\emph{inside} a single elementary cell. However, in order to
obtain the transfer matrix through the period of the structure, we
have to relate the field and its derivative at points to the left
($z_+-0$) and to the right ($z_++0$) of the boundary between two
elementary cells. Since there is no discontinuity of the
dielectric constant at this point, it might appear that all the
continuity conditions would be satisfied automatically. One should
remember, however, that functions $h_i(z)$ are different in
different elementary cells. For two adjacent $m$-th and $(m+1)$-th
cells, the respective functions, $h_i^m(z)$ and $h_i^{m+1}(z)$,
are related to each other as $h_i^m(z)=h_i^{m+1}(z-d)$, and in
order to establish connection between vectors $|c^m\rangle$ and
$|c^{m+1}\rangle$ at the boundary between these cells, one would
need to express $h_i^{m+1}(z)$ in the basis of $h_i^{m}(z)$. This
problem can be solved, but it turns out to be more convenient to
convert our transfer matrices to a more conventional basis of
plane waves
\begin{equation}\label{eq:basis_plane_wave_definition_main}
  E(z) = E_+ e^{iq_s z} + E_- e^{-iq_s z}
\end{equation}
using the conversion rule, Eq.~%
(\ref{eq:relation_transfer_matrices}), derived in the Appendix.
The resulting transfer matrix describing the evolution of the
field across an entire elementary cell  can then be presented  in
the form
\begin{equation}\label{eq:af_representation}
  T =
    \begin{pmatrix}
    af & (\bar{a}f - a\bar{f})/2 \\
    (a\bar{f} - f\bar{a})/2 & \bar{a}\bar{f}
  \end{pmatrix},
\end{equation}
which, as will be seen shortly is particularly convenient for
analysis of the spectrum of the system under consideration. This
representation is extensively used in the present paper and, in what
follows, we will refer to it as $(a,f)$-representation. The
parameters of this representation, $a$ and $f$, are defined as
\begin{equation}\label{eq:af_s_polarization}
\begin{split}
  a = g_2, \qquad f = g_1 - i S_s g_2,\\
  \bar{a} = g_2^*, \qquad \bar{f} = g_1^* + i S_s g_2^*,
\end{split}
\end{equation}
where
\begin{equation}\label{eq:g12_s_polarization}
  \begin{split}
  g_1 = \frac{1}{\sqrt{W_h}}\left[ h_1(z_+) +
  \frac{h_1'(z_+)}{iq_s}\right], \\
  g_2 = \frac{1}{\sqrt{W_h}}\left[ iq_s h_2(z_+) +
  h_2'(z_+)\right].
  \end{split}
\end{equation}

\subsubsection{$p$-polarization}

A representation similar to Eq.~(\ref{eq:af_representation}) for a
transfer matrix describing $p$-polarized light can also be
obtained along the similar lines. An important difference is that
the electric field in the $p$-polarized waves is not transverse
($\nabla\cdot\mathbf{E}\ne0$) and, therefore, satisfies a more
complex system of two differential equations. For conventional
photonic crystals the $p$-polarized waves are conveniently
described in terms of the magnetic field (see e.g.,
Ref.~\onlinecite{Joannopoulos_PC_MFL:1995}). However, the presence
of the dipole active excitations significantly reduces the
convenience of this approach because in order to close the
equation for this field, one would have to express the interaction
with excitons in terms of the magnetic field. While it can be
done, we find it more convenient to continue working with electric
field, $\mathbf{E}$.

The electric field for the $p$-polarization can be presented in the
following form:
\begin{equation}\label{eq:field_representation_p_polarization}
   \mathbf{E}(z, \boldsymbol{\rho}) =
   \left[\hat{\mathbf{e}}_\mathbf{k} E_x(z)
   + \hat{\mathbf{e}}_z E_z(z)\right]e^{i
\mathbf{k}\boldsymbol{\rho}}.
\end{equation}
Amplitudes $E_{x,z}(z)$ satisfy the system of equations
\begin{equation}\label{eq:Maxwell_equations_p_pol}
\begin{split}
 \frac{d^2E_x}{dz^2} -  i k \frac{dE_z}{dz}&
+\kappa_p^2(z)E_x = \\
-& \chi(\omega)\sum_m\Phi_m(z) \int dz\, \Phi_m(z')E_x(z'), \\
 -ik \frac{dE_x}{dz} +&
 \left[\kappa_p^2(z) - k^2\right] E_z(x) = 0,
\end{split}
\end{equation}
where $\kappa_p^2(z) = \omega^2 \epsilon(z)/c^2$. Deriving these
equations we again have explicitly taken into account that only
the in-plane component of the electric field interacts with the
heavy-hole excitons. Solving the second equation with respect to
$E_z$ we obtain the closed equation for $E_x(z)$
\begin{equation}\label{eq:E_x_equation_p_pol}
\begin{split}
  \frac{d}{dz}\left[p(z) \frac{dE_x}{dz}\right]& +
\kappa_p^2(z)E_x =
  \\
  -&\chi(\omega)\frac{4\pi\omega^2}{c^2}\sum_m\Phi_m(z)\int dz\, \Phi_m(z')E_x(z'),
\end{split}
\end{equation}
where $p(z)=\kappa_p^2(z)/[\kappa_p^2(z)-k^2]$. This function is
related to the local angle of propagation of the wave, $\theta(z)$:
$p(z) = 1/\cos\theta(z)$.

The derivation of the transfer matrix for the $p$-polarized field
follows exactly the same steps as in the previous subsection with
one important change. The Wronskian of two solutions of the
``homogeneous" (that is with $\chi\equiv 0$) version of
Eq.~(\ref{eq:E_x_equation_p_pol}) depends on the coordinate
$z$.\cite{Morse_Feshbach} This, however, does not present a major
problem, because if one takes into account that this dependence
can be presented in the form
\begin{equation}\label{eq:Wronskian_p_pol}
  W(z) = W_h\frac{p(z_+)}{p(z)},
\end{equation}
one can immediately reproduce all the main results of the previous
subsection. In particular, the transfer matrix in the basis of
plane waves can again be written down in the
$(a,f)$-representation, Eq.~(\ref{eq:af_representation}). The
parameters of this representation are given by the same
Eqs.~(\ref{eq:af_s_polarization}) and
(\ref{eq:g12_s_polarization}), where, however, one has to consider
functions $h_{1,2}$ as even and odd solutions of the
``homogeneous" version of Eq.~(\ref{eq:E_x_equation_p_pol}). Also,
the function $S_p$, which retains the same form as in
Eq.~(\ref{eq:Lorentzian_S_s_polarization}), is now characterized
by a modified radiative shift of the exciton frequency and the
decay rate. The expression for the frequency shift is obtained by
replacing the expression given by
Eq.~(\ref{eq:PC_Green_function_s_pol}) with
\begin{equation}\label{eq:PC_Green_function_p_pol}
   (G \star \mathcal{F})(z) = \frac{1}{W_h p(z_+)}\int_{z_-}^z dz'
   \mathcal{F}(z')\left[h_1(z')h_2(z)-h_1(z)h_2(z')\right]
\end{equation}
in Eq.~(\ref{eq:radiative_shift}) for the radiative shift. The rate
of exciton's decay into the $p$-polalized light is given by
\begin{equation}\label{eq:Gamma_p_def}
  \Gamma_p = \frac{2\pi\alpha \omega^2 \varphi_1^2}{q_p c^2 p(z_+)},
\end{equation}
where $q_p = \kappa_p(z_+)$. One can notice that these expressions
and corresponding expressions for the $s$-polarized wave coincide in
the case of normal propagation, i.e. when $k = 0$.

\subsection{General properties of the $(a,f)$-representation}
\label{sec:structure_transfer_matrices}

The fact that the transfer matrices for both polarizations allow
for the $(a,f)$-representation is not a coincidence but is related
to a well-known feature of the Maxwell equations. To demonstrate
it let us consider Eq.~(\ref{eq:E_x_equation_p_pol}) in a
particular case corresponding to the passive structure ($\chi
\equiv 0$). If $E(z)$ is a solution to this equation, its complex
conjugate $E^*(z)$ is also a solution. Having this in mind one can
show that that
\begin{equation}\label{eq:conservation_law_Maxwell}
  \frac{d}{dz}\left[p(z)\left(E {E^*}' - E'E^*\right)\right] =0.
\end{equation}
Now, representing the electric field in the form of Eq.~%
(\ref{eq:basis_plane_wave_definition_main}) we can find that
\begin{equation}\label{eq:conservation_law_energies}
  \frac{d}{dz}\left[q_p p(z)\left(|E_+|^2 - |E_-|^2\right)
\right]=0.
\end{equation}
This relation expresses conservation of the flux of the Poynting
vector through a plane perpendicular to the $z$-axis and
represents the fact that there are no sources or drains of the
energy in the system. The similar expression for the $s$-polarized
wave can be obtained by substitution $q_p \to q_s$, $p(z)\to1$. It
follows from Eq.~(\ref{eq:conservation_law_energies}) that the
transfer matrices through the period for both $s$ and $p$
polarizations  must preserve the combination $|E_+|^2 - |E_-|^2$,
which means that the transfer matrix written in the basis of plane
waves belongs to $SU(1,1)$ group.\cite{DNF-Modern_geometry} A
general form of an element of this group
is\cite{DNF-Modern_geometry}
\begin{equation}\label{eq:SU11_general_form}
  T=\begin{pmatrix}
    T_1 & T_2 \\
    T_2^* & T_1^*
  \end{pmatrix},
\end{equation}
where $T_{1,2}$ are complex numbers and $|T_1|^2 - |T_2|^2 = 1$.
Matrices allowing for the $(a,f)$-representation correspond to a
particular case of $SU(1,1)$ matrices, for which $T_2$ is purely
imaginary. It can be shown that such matrices describe structures
that posses a mirror symmetry. Indeed, a transfer matrix of  such
a structure satisfies the following condition $\sigma_x T\sigma_x
= T^{-1}$, where $\sigma_x$ is the Pauli matrix. Substitution of
Eq.~(\ref{eq:SU11_general_form}) into this equation gives $T_2^* =
- T_2$. The converse statement can also be verified. This property
of transfer matrices can also be proven in the general case of
matrices corresponding to Maxwell equations with external
polarization (\ref{eq:Maxwell_smooth_modulation_s_pol}) and
(\ref{eq:E_x_equation_p_pol}), but the proof is rather technical
and we do not provide it here.

Owing to the $(a,f)$-representation, the description of structures
with mirror symmetry can be significantly simplified. At the same
time, they demonstrate a rich variety of interesting phenomena,
and, therefore these structures attract a great deal of attention
(see e.g. Ref.~\onlinecite{BENDICKSON:1996}). If such a structure
is built of blocks that have mirror symmetry by themselves then
the transfer matrix through the entire structure can be written in
terms of matrix elements describing the individual blocks. Indeed,
let us consider a structure with the period $BAB$ where the
transfer matrices of the blocks $A$ and $B$ have the form
$T(a_1,f_1)$ and $T(a_2,f_2)$, respectively, in the
$(a,f)$-representation. Then it can be shown that the transfer
matrix through the period is
\begin{equation}\label{eq:simple_sandwich_transfer}
  T(a_2,f_2)T(a_1,f_1)T(a_2,f_2) = T(a,f),
\end{equation}
where
\begin{equation}\label{eq:simple_sandwich_transfer_elements}
\begin{split}
  a = a_1 a_2 f_2 - \frac{\bar{a}_1}{2}\left(\bar{a}_2f_2 -
  a_2\bar{f}_2\right) ,\\
  f = f_2 f_1 a_2 +\frac{\bar{f}_1}{2}\left(\bar{a}_2f_2 -
  a_2\bar{f}_2\right).
\end{split}
\end{equation}

We will illustrate the ``multiplication rule", Eq.~%
(\ref{eq:simple_sandwich_transfer}), considering a simple but very
important for the rest of the paper example of an elementary cell
in an optical lattice. In this system, the block $A$ corresponds
to an active element, for instance, a QW, and two blocks $B$
describe barriers, which have the \emph{same} refractive index as
block $A$. The blocks $B$ are described by a transfer matrix given
by $T_b(\phi_b) =
\mathrm{diag}\left[\exp(i\phi_b),\exp(-i\phi_b)\right]$. In this
case one has $a_2=f_2=\exp(i\phi_b/2)$ in
Eq.~(\ref{eq:simple_sandwich_transfer}) and thus
\begin{equation}\label{eq:sandwich_barriers}
T_b(\phi_b) T(a,f)T_b(\phi_b) = T(a e^{i\phi_b},fe^{i\phi_b}).
\end{equation}

If, however, one needs to take into account the difference between
refractive indices of the blocks B, and the block A, a new type of
transfer matrix products emerges. Introducing matrix $T_\rho$,
describing propagation of light across a boundary between two media
with different indexes of refraction
\begin{equation}\label{eq:interfacial_transfer_matrix}
  T_\rho = \frac{1}{1 + \rho}
  \begin{pmatrix}
    1 & \rho \\
    \rho & 1
  \end{pmatrix}.
\end{equation}
where $\rho$ is the Fresnel reflection coefficient\cite{BornWolf}
(see below Eq.~(\ref{eq:Fresnel_coefficients_vals})), we can
describe propagation of light through $BA$ and $AB$ interfaces and
across the block $A$ using the following combination of the
transfer matrices: $T_\rho(\rho)^{-1}T(a,f)T_\rho(\rho)$. An
important property of the $(a,f)$-representation is that this
product can be expressed in the following form:
\begin{equation}\label{eq:sandwich_mismatch}
T_\rho(\rho)^{-1}T(a,f)T_\rho(\rho)= \frac{1}{1-\rho^2}T(a +
\bar{a}\rho,f - \bar{f}\rho).
\end{equation}
The real factor in Eq.~(\ref{eq:sandwich_mismatch}) can be
incorporated into $a$ and $f$ due to the useful relation (with real
$\lambda$)
\begin{equation}\label{eq:af_incorporation_factor}
  \lambda T(a,f)=T(\lambda a, f) = T(a, \lambda f).
\end{equation}

Eqs.~(\ref{eq:simple_sandwich_transfer}),
(\ref{eq:sandwich_barriers}), and (\ref{eq:sandwich_mismatch})
will be extensively used throughout the paper for obtaining the
$(a,f)$-representation of various transfer matrices. This method
is often more practical than solving corresponding differential
equations. Thereby, it is interesting to note that the parameters
of the $(a,f)$-representation are essentially boundary values of
solutions of the corresponding Cauchy problem.

\section{The dispersion equation in resonant photonic crystals: the method of analysis and the band structure}

\subsection{Dispersion equation in the $(a,f)$-representation}\label{sub:disp_eq_gen}

The dispersion equation characterizing normal modes (polaritons)
in an infinite periodic one-dimensional structure can be expressed
in terms of elements of a transfer-matrix across one period of the
structure\cite{Yariv_book}
\begin{equation}\label{dispersion_law_general_definition}
  \cos Kd = \frac 1 2 \mathrm{Tr}\, T,
\end{equation}
where $K$ is the Bloch wave number, $d$ is the period of the
structure, and the transfer matrix is assumed to be written in the
plane wave basis. If this structure possesses a mirror symmetry, one
can show that the condition $\det T = 1$ can be presented as $f
\bar{a} + \bar{f} a = 2$. Using this identity we can rewrite
the polariton dispersion law in the form
\begin{equation}\label{eq:disp_law_cos}
\cos^2 \left(\frac{Kd}{2}\right) = \Re(a)\Re(f) ,
\end{equation}
where
\begin{equation}\label{eq:Re_defin}
\Re(a) = (a + \bar{a})/2.
\end{equation}

Equivalently, this dispersion equation can be presented as
\begin{equation}\label{eq:disp_law_sin}
\sin^2 \left(\frac{Kd}{2}\right) = \Im(a)\Im(f),
\end{equation}
where
\begin{equation}\label{eq:Im_defin}
\Im(a) = (a - \bar{a})/2i.
\end{equation}
If the exciton susceptibility $S$ is a real function (no
homogeneous broadening of excitons), the $\bar{a}$ coincides with
$a$ conjugated, $\bar{a}=a^*$. In this case, $\Re(a)$ and $\Im(a)$
are equivalent to regular real or imaginary parts of $a$
respectively: $\Re(a)=\mathrm{Re}(a)$ and $\Im(a)=\mathrm{Im}(a)$.
When homogeneous broadening is taken into account,
Eq.~(\ref{eq:disp_law_cos}) remains valid, but identification of
$\Re(a)$ and $\Im(a)$ with $\mathrm{Re}(a)$ and $\mathrm{Im}(a)$
can no longer be made. In this case, one has to use definitions
given in Eqs.~(\ref{eq:Re_defin}) and (\ref{eq:Im_defin}).

If the homogeneous broadening can be neglected, the r.h.s of
Eq.~(\ref{eq:disp_law_cos}) and Eq.~(\ref{eq:disp_law_sin}) are
real valued. Therefore, we can analyze the band structure in the
vicinity of the exciton resonance using the notion that for
allowed bands (real $K$) these expressions should be positive and
less than unity, while for the band gaps (complex $K$) they should
be negative or greater than unity. Thus, there are two types of
conditions determining the boundary of the bands. From
Eq.~(\ref{eq:disp_law_cos}) we have that the band boundary occurs
when either $\Re(a)\Re(f)$ or $\Re(a)\Re(f)-1$ changes sign, and
Eq.~(\ref{eq:disp_law_sin}) defines as the boundaries those
frequencies at which the change of sign occurs for expressions
$\Im(a)\Im(f)$ or $\Im(a)\Im(f)-1$. One can show that these two
pairs of conditions are equivalent to each other. More precisely,
expression $\Re(a)\Re(f)$ changes sign at the same frequencies as
expression $\Im(a)\Im(f)-1$. The same is true for the pair
$\Re(a)\Re(f)-1$ and $\Im(a)\Im(f)$. Thus, all the band boundaries
can be found by considering changes of sign of expressions
$\Re(a)\Re(f)$ and $\Im(a)\Im(f)$, with negative values
corresponding to the band gaps. The factorized form of right-hand
sides of the polariton dispersion equation,
Eqs.~(\ref{eq:disp_law_cos}) and (\ref{eq:disp_law_sin}),
drastically simplifies the analysis of the spectrum, as it will be
seen in the subsequent sub-sections.

The factorization of the dispersion equations,
Eqs.~(\ref{eq:disp_law_cos}) and (\ref{eq:disp_law_sin}) becomes
possible only owing to the $(a,f)$-representation of the
transfer-matrix, which makes this representation particularly
suitable for studying the band structure and dispersion laws of
polaritons in resonant photonic crystals. At the same time, this
representation is not very convenient for calculating
relfection/transmission spectra of finite structures because of a
cumbersome relation between the parameters of a single layer
transfer matrix and a matrix describing the entire structure. This
problem, however, can be conveniently solved with the help of a
different representation introduced in our recent paper,
Ref.~\onlinecite{Optical_Properties_MQWPC}. It is useful to
establish a direct relationship between these two representations.
First we notice that an arbitrary matrix of the form
(\ref{eq:af_representation}) can be written in the form of a
transfer matrix describing propagation of a wave across a single QW
[see Eq.~(\ref{eq:transfer_matrix_well}) below]. This can be done by
introducing an effective optical width of one period of the
structure, $\tilde{\phi}$, and an effective excitonic
susceptibility, $\tilde{S}$. The relation between the parameters of
the $(a,f)$-representation and these effective parameters can be
found by comparing Eqs. (\ref{eq:af_representation}) and
(\ref{eq:transfer_matrix_well})
\begin{equation}\label{eq:relation_af_effective}
  \tilde{S} = \frac{1}{2i} (a \bar{f} - \bar{a} f), \qquad
  e^{i\tilde{\phi}} = \frac{a}{\bar{a}}.
\end{equation}
The usefulness of this relation is twofold. First, it shows that
for any resonant photonic crystal the transfer matrix through the
period can be represented as a transfer matrix through a
conventional MQW structure. Second,
Eq.~(\ref{eq:relation_af_effective}) can be used for a
consideration of reflection and transmission spectra of a
structure described by a transfer matrix in
$(a,f)$-representation. This is where the introduction of
$\tilde{S}$ and $\tilde{\phi}$ and then, consecutively, the
representation the transfer matrix in terms of parameters $\theta$
and $\beta$ of Ref.~\onlinecite{Optical_Properties_MQWPC} becomes
very convenient. However, this representation misses the
possibility to factorize the dispersion equation, so it is not
very suitable for a constructive analysis of the dispersion law.
One can see, therefore, that the representations of the transfer
matrix in terms of parameters $(a,f)$ and  $\tilde{S}$ and
$\tilde{\phi}$ compliment each other in a sense that each of them
is suited best for its own set of problems.

\subsection{MQW structures with a simple elementary cell}

\begin{figure}
  \includegraphics[width=3in]{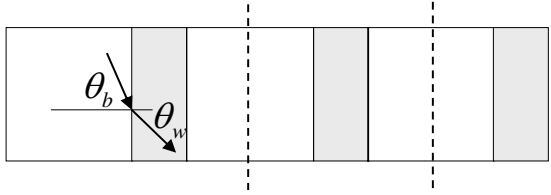}\\
  \caption{The periodic structure built of two blocks. Vertical
  dashed lines show the boundary of the elementary cell. The angles
  of propagation inside the blocks are related by
  Snell's law.}\label{fig:two_blocks_elementary_cell}
\end{figure}

As it was mentioned in Introduction, MQW structures present one
example of RPCs, in which effects due to reflection of light from
well-barrier interfaces coexist with effects due to the resonant
light-exciton coupling. The spatial profile of the refractive
index in these structures is particulary simple, and is described
by a piece-wise constant function
[Fig.(\ref{fig:two_blocks_elementary_cell})]. Various properties
of these structures have been studied in a number of
publications\cite{IvchenkoContrast,JointComplex,Kavokin_oblique,AndreaniPRB1992,time_resolved_contrast,DielEnv,Optical_Properties_MQWPC},
including papers
Refs.\onlinecite{IvchenkoContrast,JointComplex,Optical_Properties_MQWPC}
dealing particularly with dispersion laws and the band structure
of their normal modes.
Nevertheless, we find it useful to consider this case in details
in this paper because it gives a clear illustration of using the
$(a,f)$-representation for the analysis of complex dispersion
equations.

We start by considering two particular cases: a passive photonic
crystal (no resonances) and an optical lattice (no refractive index
mismatch). This will give us useful benchmarks  for discussing the
general situation. In the latter case we will neglect the exciton
homogeneous broadening, which allows us to identify operations $\Re$
and $\Im$ with taking regular real and imaginary parts of the
respective expressions. The same is obviously true for the case of
passive structures.

\subsubsection{Passive photonic crystals}
Let us consider a structure built of a periodic sequence of two
blocks characterized by the widths $d_{b,w}$ and the indices of
refraction $n_{b,w}$. To emphasize the mirror symmetry we choose the
elementary cell as shown in
Fig.~\ref{fig:two_blocks_elementary_cell}. The transfer matrix
through the period of the structure has the form
\begin{equation}\label{eq:transfer_matrix_period_PC}
  T = T_b^{1/2}T_\rho^{-1}T_w T_\rho T_b^{1/2},
\end{equation}
where
\begin{equation}\label{eq:transfer_matrix_barrier}
  T_{b,w} =
  \begin{pmatrix}
    e^{i \phi_{b,w}} & 0 \\
    0 & e^{-i \phi_{b,w}}
  \end{pmatrix},
\end{equation}
$\phi_{b,w} = \omega n_{b,w} d_{b,w} \cos\theta_{b,w}/c$ and
$\theta_{b,w}$ are the angles of propagation of the wave.

The scattering of the electromagnetic wave at the interface between
different blocks depends on both the angle of incidence of the wave
and its polarization state. These effects are described by Fresnel
coefficients $\rho_s$ and $\rho_p$,
\begin{eqnarray}\label{eq:Fresnel_coefficients_vals}
 \rho_s = \frac{n_w \cos\theta_w - n_b \cos\theta_b}{n_w
\cos\theta_w + n_b
 \cos\theta_b},
 \nonumber \\
 \rho_p = \frac{n_w \cos\theta_b - n_b \cos\theta_w}{n_w
\cos\theta_b + n_b
 \cos\theta_w},
\end{eqnarray}
for $s$ and $p$ polarizations, respectively. Below we denote the
Fresnel coefficients simply by $\rho$ having in mind that for a
particular polarization only one of these expressions should be
used.

Using Eqs.~(\ref{eq:sandwich_barriers}) and
(\ref{eq:sandwich_mismatch}) we can find parameters of the
$(a,f)$-representation of the transfer matrix, which in this case
take the following form 
\begin{equation}\label{eq:af_parameters_passive_multilayer}
  a = \frac{e^{i\omega\tau_+} + \rho e^{i\omega \tau_-}}{1-\rho^2}, \qquad
  f = e^{i\omega\tau_+} - \rho e^{i\omega \tau_-},
\end{equation}
where $\tau_\pm =(\phi_b \pm \phi_w)/2\omega$. Now we can use
general dispersion equations, Eq.~(\ref{eq:disp_law_cos}) and
(\ref{eq:disp_law_sin}) to analyze the structure of the allowed
and forbidden bands of this system. As we discussed in
Section~\ref{sub:disp_eq_gen} there are two types of forbidden
bands, determined by conditions $\mathrm{Re}\,(a)\mathrm{Re}\,(f)
< 0$, and, $\mathrm{Im}\,(a)\mathrm{Im}\,(f) < 0$ respectively. If
$\tau_+$ and $\tau_-$ are incommensurate, these band gaps
alternate along the frequency axis and  can be classified as
either odd (first, third, etc.) or even (second, forth, etc.).

Since the coefficients of the dispersion equation in the passive
photonic crystals do not contain any singularities, its right hand
side can only change the sign by passing through zero. Thus, the
band boundaries are situated at the frequencies where either
$\mathrm{Re}\,(a)\mathrm{Re}\,(f) =0$ or
$\mathrm{Im}\,(a)\mathrm{Im}\,(f) = 0$. For instance, if $\rho>0$,
the boundaries of odd band gaps are determined by equations
\begin{equation}\label{eq:gap_edges_PC}
 \mathrm{Re}\,[a(\Omega_+)]= 0, \qquad \mathrm{Re}\,[f(\Omega_-)]=
 0,
\end{equation}
where $\Omega_-$ and $\Omega_+$ correspond to the lower and higher
frequency boundaries of a given band gap respectively. The
explicit form of these equations can be obtained using
definitions, Eq.~(\ref{eq:af_parameters_passive_multilayer}):
\begin{equation}\label{eq:gap_edges_PC_explicitely}
  \begin{split}
  \cos (\Omega_-\tau_+) - \rho \cos (\Omega_-\tau_-) = 0,\\
 \cos (\Omega_+\tau_+) + \rho \cos (\Omega_+\tau_-) = 0.
  \end{split}
\end{equation}
In the simplest case, when the layers have the same optical width,
one has $\tau_- = 0$ and the positions of the edges of the
forbidden gap are $\omega_r(1\pm 2 \arcsin(\rho)/\pi)$, where
\begin{equation}\label{eq:omega_r_definition}
  \omega_r\tau_+ = \frac{\pi}{2}.
\end{equation}
While this case gives a convenient reference point, however,
having in mind applications to MQW structures, an opposite
situation, when the optical widths of the layers are different, is
of more interest. Generally, as one can see from
Eq.~(\ref{eq:gap_edges_PC_explicitely}), the boundaries of the gap
are situated asymmetrically with respect to $\omega_r$. Under the
assumption of narrow gaps, which is fulfilled for a not very
strong contrast of the refraction indexes,  and for angles not too
close to the angle of the total internal reflection, the positions
of the boundaries are given by
\begin{equation}\label{eq:gap_edges_PC_solution}
  \Omega_\pm = \omega_r \left(1\pm \frac{2\rho \sin\phi_w}{\pi
(1\pm \rho
  \cos\phi_w)}\right).
\end{equation}

\subsubsection{Semiconductor optical lattice}

In this case we assume that all layers in the structure have the
same index of refraction, $n$, but there are dipole active
excitations in QWs. The propagation of light through a QW is
described by the transfer matrix of the form
\begin{equation}\label{eq:transfer_matrix_well}
  T_w = \begin{pmatrix}
    e^{i\phi_w}(1-i S) & -i S \\
    i S & e^{-i\phi_w}(1 + i S)
  \end{pmatrix},
\end{equation}
where $\phi_w = \omega n d_w\cos \theta_w/c$. The parameters of the
$(a,f)$-representation of $T_w$ are
\begin{equation}\label{eq:a_f_parameters_quantum_well}
  a = e^{i\phi_w/2}, \qquad f = e^{i\phi_w/2}(1-iS).
\end{equation}

The excitonic contribution to the scattering of light is described
by
\begin{equation}\label{eq:S_def_simple}
  S = \frac{\Gamma_0}{\omega - \omega_0 + i\gamma}.
\end{equation}
The radiative decay rate, $\Gamma_0$, depends on the angle of
incidence. As follows from Eqs.~(\ref{eq:Gamma_s_def}) and
(\ref{eq:Gamma_p_def}), for structures with homogeneous dielectric
function these dependencies for different polarizations are
\cite{Andreani_SSC,Citrin_Lifetime,IvchenkoContrast}
\begin{equation}\label{eq:Gamma_sp_oblique}
  \Gamma_0^{(s)} = \Gamma_0 /\cos \theta_w,
  \qquad \Gamma_0^{(p)} = \Gamma_0 \cos \theta_w.
\end{equation}

The transfer matrix through the period of the structure is
\begin{equation}\label{eq:transfer_matrix_period_MQW}
 T = T_b^{1/2}T_w T_b^{1/2}.
\end{equation}
Taking into account the multiplication rule
(\ref{eq:sandwich_barriers}), we can find parameters of the
$(a,f)$-representation for the entire matrix
(\ref{eq:transfer_matrix_period_MQW}) and the respective
dispersion equation\cite{Keldysh,IvchenkoSpectrum,CitrinSpectrum}
\begin{equation}\label{eq:optical_lattice}
\cos^2\left(\frac{Kd}{2}\right) =
\cos\omega\tau_+(\cos\omega\tau_+ + S \sin\omega\tau_+).
\end{equation}
Unlike the case of the passive photonic crystal, the coefficients
of the $(a,f)$-representation now contain a singularity at the
frequency of the exciton resonance (we neglect the homogeneous
broadening of the excitons here). Therefore, the r.h.s. of the
dispersion equation (\ref{eq:optical_lattice}) can change its sign
by passing not only through zero, but also through infinity. The
latter happens at $\omega_0$, which becomes one of the boundaries
of the band gap associated with the exciton resonance. In general
case, the width of this and all other allowed and forbidden bands
are proportional to $\Gamma$, which is many orders of magnitude
smaller than $\omega_0$. Therefore, usually, exciton related
modifications of the photon dispersion law in the optical lattice
are negligible. This situation changes, however, if we require
that this singularity is canceled by the term $\cos
(\omega\tau_+)$, which happens if $\cos (\omega_0 \tau_+) = 0$.
This is equivalent to the condition for the Bragg resonance, when
the half-wavelength of the radiation at the exciton frequency is
equal to an odd multiplier of the period of the structure
\begin{equation}\label{eq:standard_Bragg_condition}
  \omega_0 \tau_+ = \frac{\pi}{2}+\pi n.
\end{equation}
The spectrum of such structures is characterized by a much wider
band gap with the width equal to
\begin{equation}\label{eq:gap_width_MQW}
  \Delta_\Gamma = 2\sqrt{\frac{\Gamma_0}{\tau_+}} = 2
  \sqrt{\frac{2\Gamma_0\omega_0}{\pi(1+2n)}}
\end{equation}
and is indicative of enhanced coupling between light and QW
excitons.

\subsubsection{MQW structures with a mismatch of the
indices of refraction} \label{sec:MQW_mismatch}

Now we are ready to discuss the general case of MQW structures
with the contrast of the refractive indices. The transfer matrix
for this structure is obtained from Eq.~%
(\ref{eq:transfer_matrix_period_MQW}) by taking into account the
scattering at the interfaces between the QWs and the barriers,
\begin{equation}\label{eq:transfer_matrix_period_PCMQW}
 T = T_b^{1/2}T_\rho^{-1} T_w T_\rho T_b^{1/2}.
\end{equation}
The dispersion equation following from the $(a,f)$-representation of
this transfer matrix has the form
\begin{equation}\label{eq:disp_law_cos_mismatch}
\cos^2 \left(\frac{Kd}{2}\right) =
\frac{1}{1-\rho^2}\mathrm{Re}(a_{PC})[\mathrm{Re}(f_{PC}) + S\,
\mathrm{Im}(a_{PC})],
\end{equation}
where $a_{PC}$ and $f_{PC}$ are calculated for a passive
multilayer structure and are given by
Eqs.~(\ref{eq:af_parameters_passive_multilayer}).

\begin{figure}
  \includegraphics[width=3in]{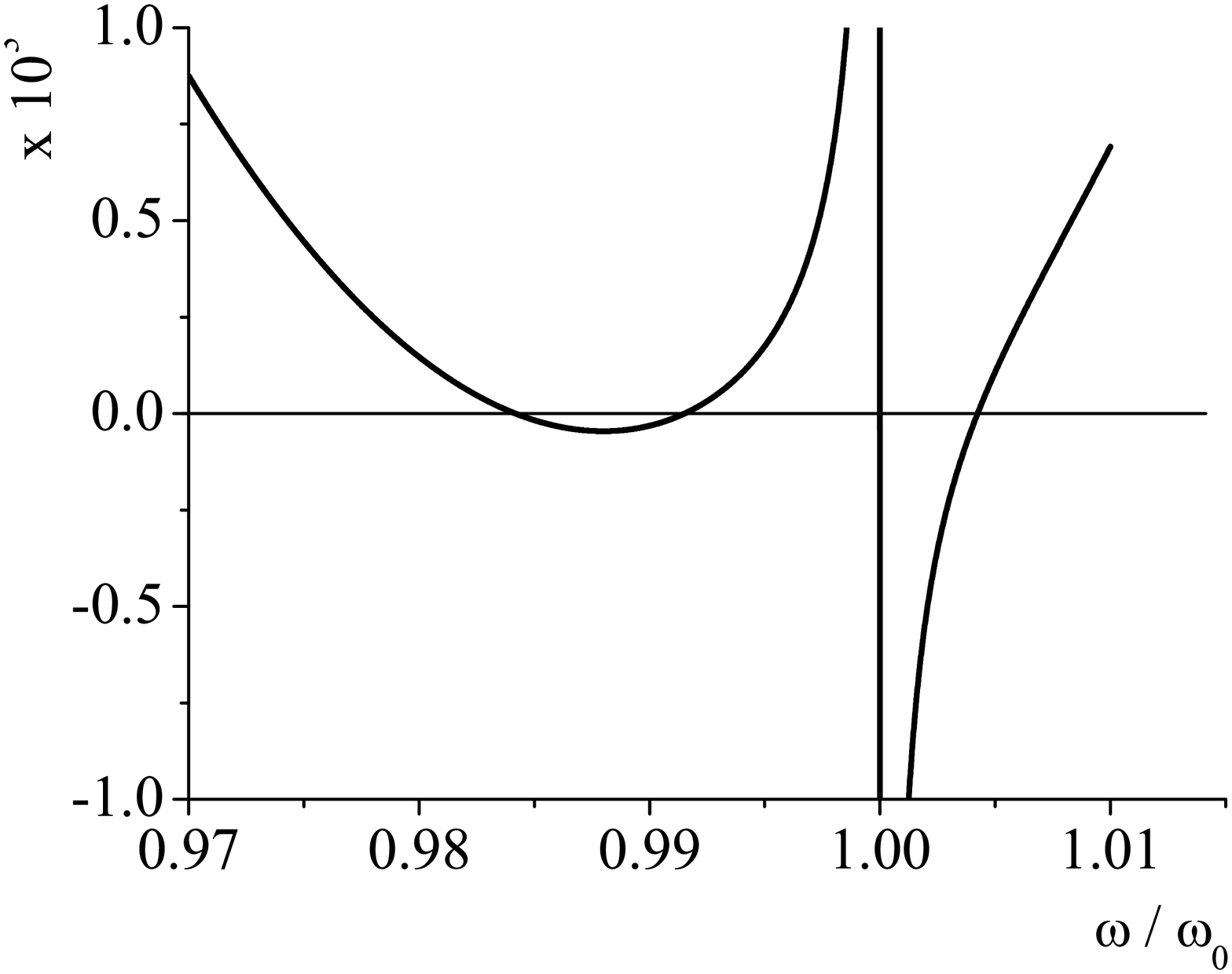}\\
  \caption{The figure plots the r.h.s. of Eq.~(\ref{eq:disp_law_cos_mismatch}) scaled by $10^3$
  for a structure slightly detuned from the Bragg resonance as a function of frequency. The material
parameters are chosen to be close to typical parameters of
GaAs/Al${}_x$Ga${}_{1-x}As$ structures: $\rho = 0.03$, $\Gamma_0 =
60$ $\mu$eV, $\omega_0 = 1.5$ eV. The frequencies where this
expression is negative or exceeds unity (not shown in this plot)
correspond to band gaps. The vertical line shows the position of the
exciton frequency.}\label{fig:structure_of_MQWPC_gap}
\end{figure}

Similarly to the case of the optical lattice, the structure of the
spectrum near the exciton frequency is complicated by the singular
character of the excitonic susceptibility (see
Fig.~\ref{fig:structure_of_MQWPC_gap}), which gives rise to new
branches of exciton related collective
excitations\cite{Raikh_braggaritons} and respective band gaps. We,
however, will again focus on Bragg structures, characterized by
the cancellation of the excitonic singularity. As a result, an
anomalously broad band gap in the vicinity of the exciton
frequency is formed. As one can see from
Eq.~(\ref{eq:disp_law_sin}), such a cancelation occurs when either
$\mathrm{Re}(a_{PC}(\omega_0))=0$ or
$\mathrm{Im}(a_{PC}(\omega_0))=0$. Both these cases imply that the
exciton frequency coincides with a boundary of one of the photonic
band gaps. Most experiments with these systems tend to deal with
structures having as short period as possible. Therefore, we only
consider the case, when $\omega_0=\Omega_+$, where $\Omega_+$ is
the boundary of the lowest band gap of the respective photonic
crystal, determined by the first of equations
(\ref{eq:gap_edges_PC}). The exciton related band gap in this case
appears at the frequencies where the r.h.s. of
Eq.~(\ref{eq:disp_law_cos_mismatch}) is negative.

Assuming a smallness of the gap we can expand $a_{PC}$ and
$f_{PC}$ near the frequencies $\Omega_{\pm}$
\begin{equation}\label{eq:apprs_a_b}
  \mathrm{Re}[a_{PC}(\omega)] = (\Omega_+ - \omega)t_+,
\quad
  \mathrm{Re}[f_{PC}(\omega)] = (\Omega_- - \omega)t_-,
\end{equation}
and write the equation for the boundaries of the forbidden gaps in
the form
\begin{equation}\label{eq:Bragg_gap_edges_eqs_mismatch}
 (\omega - \Omega_+)(\omega - \Omega_-) -
 \Gamma_0\frac{\mathrm{Im}(a_{PC})}{t_-}= 0,
\end{equation}
where
\begin{equation}\label{eq:T_pm_expressions}
 t_+ =
\left|\frac{d}{d\omega}\mathrm{Re}\,[a_{PC}(\omega)]\right|_{\Omega_+},
\quad
 t_- =\left|\frac{d}{d\omega}\mathrm{Re}[f_{PC}(\omega)]\right|_{\Omega_-}.
\end{equation}
In Eqs.~(\ref{eq:apprs_a_b}) we explicitly have taken into account
the negative sign of these derivatives. In
Eq.~(\ref{eq:Bragg_gap_edges_eqs_mismatch}) the radiative decay
rate, $\Gamma_0$, is defined in Eqs.~(\ref{eq:Gamma_sp_oblique})
should be taken according to the polarization of the wave and the
angle of propagation.

Using these approximations we can present the boundaries of the
polariton band gap as
\begin{equation}\label{eq:Bragg_gap_edges_sols_mismatch}
  \omega_\pm  = \Omega_c \pm \frac{1}{2} \Delta,
\end{equation}
where $\Omega_c=(\Omega_-+\Omega_+)/2$ is the center of the gap
and $\Delta$ is its width. The expression for $\Delta$ can be
written in the following form:
\begin{equation}\label{eq:Bragg_gap_widths}
  \Delta = \sqrt{\Delta_{PC}^2 + \tilde{\Delta}_\Gamma^2}
\end{equation}
and is equal to ``Pythagorean sum" of the widths of the passive
photonic gap, $\Delta_{PC} = |\Omega_+ - \Omega_-|$, and the
modified excitonic gap,
\begin{equation}\label{eq:PCMQW_effective_Delta}
  \tilde{\Delta}_\Gamma^2 =
\Gamma_0\frac{4\mathrm{Im}(a_{PC})}{t_-}
  \approx \Delta_\Gamma^2\frac{1+\rho}{1-\rho}.
\end{equation}
The last formula in this equation is obtained by using approximate
expressions $\mathrm{Im}(a_{PC}) \approx 1 + \rho$ and $t_- \approx
\tau_+ ( 1 - \rho)$ which are valid provided that the optical width of the wells is very
small compared to the width of the barriers.

The equation
\begin{equation}\label{eq:Bragg_generalized}
\omega_0=\Omega_+
\end{equation}
generalizes the condition for the Bragg resonance given by
Eq.~(\ref{eq:standard_Bragg_condition}) for optical lattices.
Indeed, normal modes in photonic crystals are characterized by the
Bloch wave number $K_{PC}$, rather than by the wave number of a
homogeneous medium $\omega n/c$. For the band boundaries
$\omega=\Omega_\pm$ one has $K_{PC}(\Omega_\pm) d=\pi$, so that
$\omega_0=\Omega_+$ is equivalent to $K_{PC}(\omega_0)d=\pi$, which
is a direct generalization of
Eq.~(\ref{eq:standard_Bragg_condition}), expressed in terms of the
appropriate wave number.

While the results presented have been obtained using the
assumption that $\rho >0$ (that is, for the normal propagation,
$n_w > n_b$) they remain valid in the opposite case due to the
symmetry of the transfer matrix under the transformations $\rho\to
-\rho$ and $a_{PC} \leftrightarrow f_{PC}$. In other words, when
we have the opposite relation between $n_w$ and $n_b$ all the
arguments used above can be repeated with the mirror reflection of
the frequency axis with respect to the center of the photonic gap.
In particular, the Bragg resonance occurs when the exciton
frequency coincides with the left (low frequency) edge of the
photonic band gap.

\subsubsection{Angular dependence of the band structure}

In previous publications on the band structure of MQW
systems\cite{JointComplex} only modes propagating along the growth
direction of the structure were considered. In this paper, thanks to
a general nature of our approach, we can consider waves propagating
at an arbitrary angle treating both $s$- and $p$- polarizations of
the electromagnetic waves on equal footing. The foundation for this
consideration is laid by the results of the previous subsections of
the paper, where the expressions for the parameters defining the
band structure were derived in terms of Fresnel coefficients, Eqs.
(\ref{eq:Fresnel_coefficients_vals}). The latter contains all the
information about angular and polarization dependencies of resonant
as well as refractive index contrast contributions to the band
structure.
\begin{figure}
  \includegraphics[width=3in]{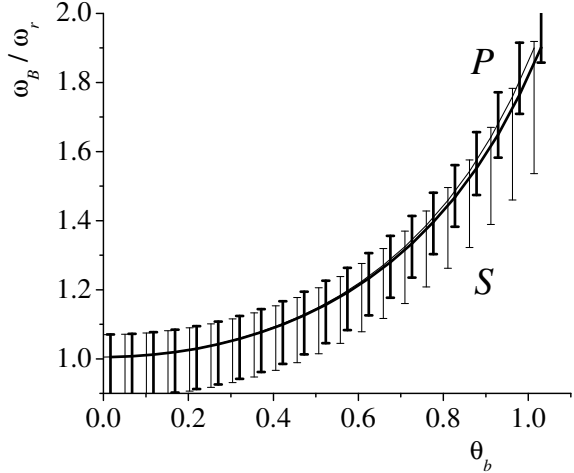}\\
  \caption{Dependence of the Bragg resonance frequency
  $\omega_B/\omega_r$ [see Eq.~(\ref{eq:omega_r_definition})] on the
  angle of propagation $\theta_b$ measured in the barriers. The material parameters
  are the same as in Fig.~\ref{fig:structure_of_MQWPC_gap}. Bold and
  thin lines correspond to $p-$ and $s$-polarizations, respectively.
  The vertical lines (the error bars) show the forbidden gap for
  each polarization. For better visibility the gap is scaled by the
  factor of $5$.
  }\label{fig:Gap_Oblique_Bragg}
\end{figure}

It is clear from Eq.~(\ref{eq:gap_edges_PC_solution}) that the
Bragg frequency defined by the generalized Bragg condition,
Eq.~(\ref{eq:Bragg_generalized}), depends on the propagation angle
of the wave. This dependence is presented in
Fig.~\ref{fig:Gap_Oblique_Bragg} for both $s$- and $p$-polarized
modes. Due to the narrowness of the QWs the position of the Bragg
resonance approximately follows the renormalization of the optical
width of the period of the structure $\propto\cos(\theta_b)$ for
both polarizations. This effect can be used to tune the position
of the band gap of the structure. Indeed, if one changes the
position of the exciton frequency, by, for instance, using quantum
confined Stark effect,\cite{SCHMITT-RINK:1989} the system can be
tuned back to the Bragg resonance by changing the propagation
angle of the wave. Assuming that the structure remains tuned to
the Bragg condition with the changing angle we can consider the
angular dependence of the width of the polariton band gap. One can
see from Fig.~\ref{fig:Gap_Oblique_Bragg}, where this width is
presented by vertical bars, that its angular dependence is
different for different polarizations. For the $s$-polarization
the width monotonously increases when the angle increases, while
for the $p$-polarization it, first, decreases, reaches its minimum
at the Brewster's angle determined by the equation $\sin\theta_b =
n_w/\sqrt{n_b^2 + n_w^2}$ (it corresponds to $\theta_b \approx
0.8$ in Fig.~\ref{fig:Gap_Oblique_Bragg}), where the only
contribution to the gap is due to the exciton-light interaction,
and then starts increasing. It should be noted that at the
Brewster's angle the Fresnel coefficient $\rho_p$ changes its
sign. Thus, as it has been pointed above, the condition of the
Bragg resonance for the $p$-polarized wave propagating at angles
larger than the Brewster's angle is formulated as equality of the
exciton frequency and the lower frequency edge of the photonic
band gap for the $p$-polarized wave.

Structures with the fixed exciton frequency and the period can  be
tuned to the Bragg resonance at a single angle only. For waves
propagating at other angles the Bragg condition is violated, and
the structure becomes detuned from the resonance. In the case of
small detunings, it is natural to refer to such systems as
off-Bragg systems. Thus, the discussion of the angular dependence
of the band structure, naturally involves considering properties
of the off-Bragg structures.

When the exciton frequency is not tuned to $\Omega_+$, one has to
take into account the singularity of the exciton susceptibility
when looking for the boundaries between allowed and forbidden
bands. This singularity results in a narrow band gap situated
between zeros of Eq.~(\ref{eq:disp_law_sin}), i.e. where
$\sin^2(Kd/2) < 0$. This band gap occupies the region between
$\omega_0$ and $\omega_0 + \Omega_\delta$, where
\begin{equation}\label{eq:omega_delta}
\Omega_\delta = \frac{\pi}{2}\tau_+ \delta
\tilde{\Delta}_\Gamma^2
\end{equation}
and $\delta = \omega_0 - \Omega_+ \ne 0$. For slightly off-Bragg
structures, the width of this addition to the band gap is small,
and we will neglect its existence in the future discussions. Thus,
the band boundaries are determined by zeros of r.h.s.~of
Eq.~(\ref{eq:disp_law_cos}), and as the result the band structure
is determined by four frequencies: $\omega_0$ and the roots of the
equation
\begin{equation}\label{eq:Bragg_off_gap}
(\omega - \Omega_+)\left[(\omega - \Omega_-)(\omega-\omega_0)   -
\frac{\tilde{\Delta}^2_\Gamma}{4}\right] = 0.
\end{equation}

The band structure is characterized by two band gaps and one
allowed band between them. If we put these four frequencies in
ascending order, the band gaps will lie between the first and
second pairs of frequencies separated by a transparency window
(see Fig.~\ref{fig:structure_of_MQWPC_gap}). The exact order of
the band boundaries depends on relation between $\omega_0$ and
$\Omega_+$. If $\omega_0 < \Omega_+$ then the band gaps are
between $\omega'_-$ and $\omega_0$ and between $\Omega_+$ and
$\omega'_+$, where
\begin{equation}\label{eq:sec_II_roots_gap}
 \omega'_\pm =
 \Omega_c +\frac{\delta}{2} \pm  \frac{1}{2} \sqrt{(\Delta_{PC} -
 \delta)^2+\tilde{\Delta}_\Gamma^2},
\end{equation}
while the allowed band is between $\omega_0$ and $\Omega_+$. Thus,
detuning the exciton resonance frequency away from $\Omega_+$
leads to the appearance of the transparency window between
$\omega_0$ and $\Omega_+$ inside the band gap obtained for the
Bragg case $\omega_0 = \Omega_+$, and to a slight modification of
the external edges of the gap. This result is in a qualitative
agreement with the earlier analysis of the off-resonant MQW
structures.\cite{DLPRBSpectrum} The transparency window would
manifest itself in optical spectra as a dip in the reflection
coefficient. A similar dip was actually observed in Refs.~%
\onlinecite{KhitrovaMQWprl,KhitrovaMQWprb,Optical_Properties_MQWPC},
where structures with the period satisfying the non-modified Bragg
condition, Eq.~(\ref{eq:standard_Bragg_condition}), were
considered. Since the well and barrier materials had different,
albeit close, values of the refractive indexes, the real Bragg
period should have been determined from
Eq.~(\ref{eq:Bragg_generalized}), and the structures considered in
those papers were actually slightly off-Bragg. It should be
mentioned, however, that the detuning from the real Bragg
condition could be not the only cause for the observed dip in
reflection. The inhomogeneous broadening of the excitons could
also contribute to this effect.\cite{OmegaDefectPRB}

\begin{figure}
  \includegraphics[width=3in]{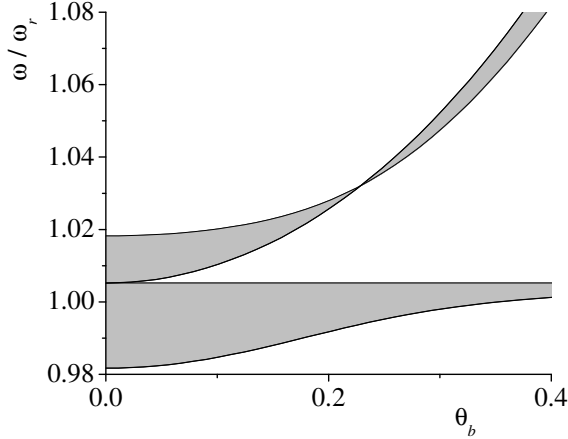}\\
  \caption{The dependence of the band gap structure of the
$s$-polarized wave on the angle of
  propagation $\theta_b$ measured in the barriers. The material parameters
  are the same as in Fig.~\ref{fig:structure_of_MQWPC_gap}. The dashed
regions correspond
  to the forbidden gaps. The structure is assumed to be tuned
  to the Bragg resonance at normal
propagation.}\label{fig:Gap_angle}
\end{figure}

Now we can describe how the band structure of our system evolves
with the changing propagation angle. Let us assume that the system
is at the Bragg resonance for the normal incidence. Increasing the
angle, we detune it away from the resonance. As a result, two band
gaps, one adjacent to the exciton frequency and the other one
detached from it, separated by the transparency window, appear.
Fig.~\ref{fig:Gap_angle} shows the angular dependence of the band
structure for the case of the $s$-polarization. An interesting
phenomenon is seen to occur at an angle at which $\Omega_+ =
\omega_+'$. In this case, the width of the band gap detached from
the exciton frequency turns to zero (band gap collapse). This
collapse is a specific feature of RPCs and is absent in both
purely passive structures and optical lattices. An important
property of the gap adjacent to $\omega_0$ is its weak
polarization dependence. This can be understood if one notices
that any difference between polarizations could only appear at
angles where $\omega_B - \omega_0 \gg \tilde\Delta_\Gamma,
\Delta_{PC}$. The width of the adjacent gap for these angles can
be found from Eq.~(\ref{eq:sec_II_roots_gap}). It is equal to
\begin{equation}\label{eq:adjacent_gap_big_angles}
  \Delta_{\textrm{adj}} \approx
\frac{\tilde\Delta_\Gamma^2}{\Omega_- - \omega_0}
\end{equation}
With the accuracy up to the terms $\sim \Delta_{\textrm{adj}}
\Delta_{PC}/(\omega_B - \omega_0)$, it is the same for both
polarizations. The property of the omnidirectional reflectivity,
i.e. a resonant reflection at all angles of incidence, ensues from
these results. It follows from the notion that a usual scattering
problem is set up for a structure embedded in a medium with
essentially lower index of refraction (air or vacuum). As the
result, the angles of propagation inside the structure cannot
exceed the angle of the internal reflection $\theta_c$ at the
boundary between the structure and surrounding medium. Therefore,
for all angles of incidence the range of frequencies between
$\omega_0$ and $\omega_-'(\theta_c)$ corresponds to the forbidden
gap and, therefore, to a resonant reflection. The similar effect
has been considered for the case of passive photonic crystals in a
number of
publications.\cite{WINN:1998,CHIGRIN:1999,BRUYANT:2003,YONTE:2004}
Here we want to emphasize the feature specific for the RPC. If the
angle of total internal reflection is small (for the air-GaAs
interface $\theta_c\approx 0.28$), we can describe the change of
the edges of the photonic band gap by a simple renormalization of
the optical width of the period, $\Omega_\pm \rightarrow
\Omega_\pm/\cos\theta$. It results in the width of the region
corresponding to the omnidirectional reflection in the form
\begin{equation}\label{eq:omnidirectional_width}
  \Delta_{\mathrm{omni}} \approx \frac{1}{2}\,\frac{\tilde{\Delta}_\Gamma^2}{
  \Omega_+ \sin^2 \theta_c - 2\Delta_{PC}},
\end{equation}
where we have assumed that the photonic forbidden gap is not too
wide, i.e. $\Omega_+ \sin^2 \theta_c
> 2\Delta_{PC}$. One can see that the presence of QWs
essentially weakens the condition of the omnidirectionality in
comparison with passive photonic crystals.

\subsection{General modulation of the dielectric function}
\subsubsection{Resonant photonic crystals}

\begin{figure}
\includegraphics[width=3.2in]{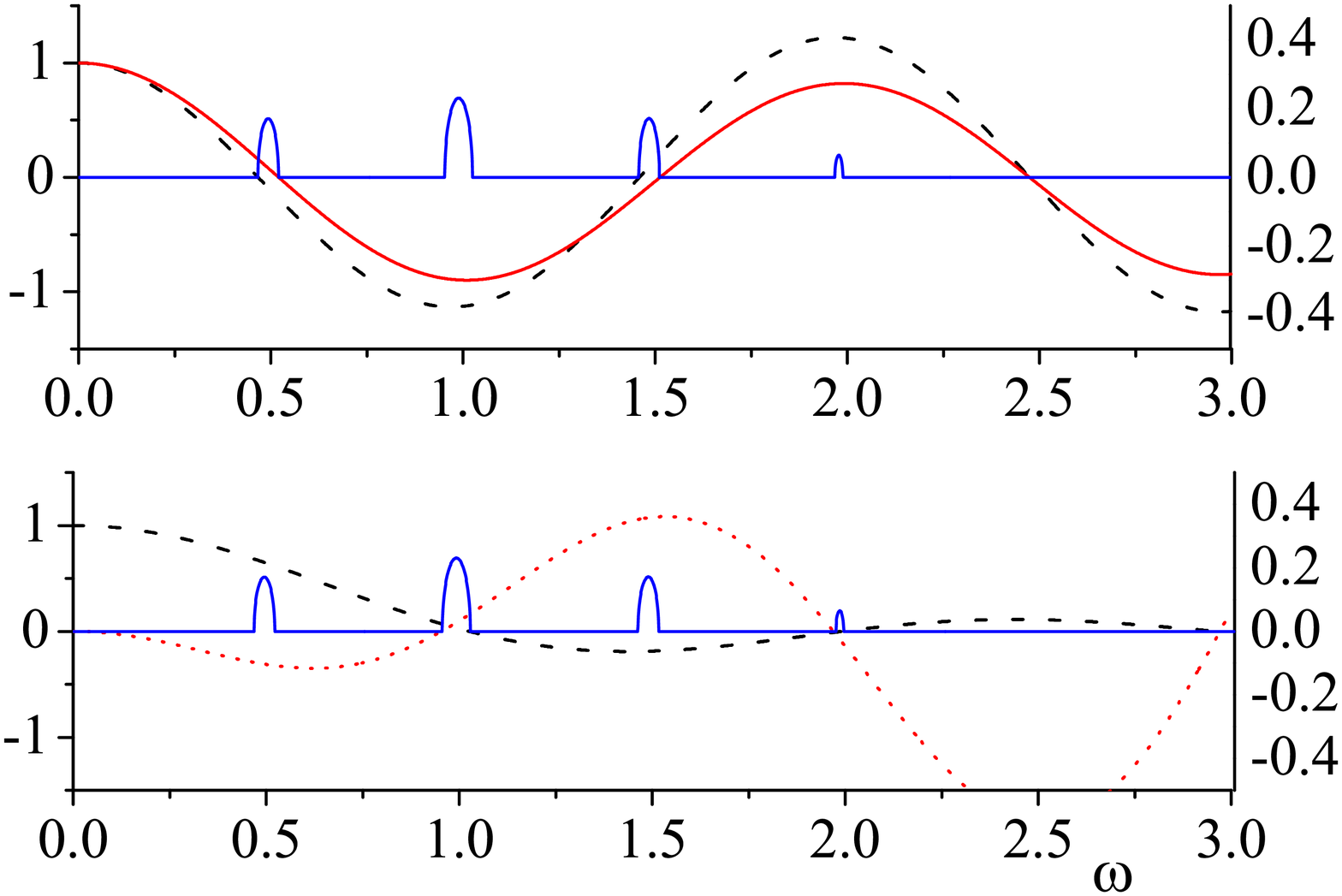}\\
\caption{Dependence of $\mathrm{Im}(Kd)$  (solid lines, left axes)
and wave functions and its derivative (dashed and dotted lines,
respectively, right axes) on the frequency for a passive structure
($S\equiv0$). The non-zero values of $\mathrm{Im}(Kd)$ correspond
to the band gaps. The index of refraction is $n(z) = 3 +
\cos^{20}(\pi z/2 )$. (a) $Kd$ is the solution of
Eq.~(\ref{eq:dispersion_law_smooth_modulation}),  $h_1(\omega)$,
the dotted line depicts $h_2'(\omega)$. (b) $Kd$ has been found
from Eq.~(\ref{eq:dispersion_law_smooth_modulation_sin}), the
dashed line represents $h_2(\omega)$, the dotted line depicts
$h_1'(\omega)/5$. One can see how zeroes of the wave function and
its derivative determine consecutive band
boundaries.}\label{fig:gaps_oscillating_theorem}
\end{figure}

The formalism developed in this paper allows us to analyze the
band structure in RPCs with an arbitrary periodic modulation of
the dielectric function.  Using the $(a,f)$-representation for
transfer matrices and expressions for the parameters $a$ and $f$
given by Eqs.~(\ref{eq:af_s_polarization}), we can derive the
dispersion equation for such a structure in the form similar to
Eq.~(\ref{eq:disp_law_cos}):
\begin{equation}\label{eq:dispersion_law_smooth_modulation}
\cos^2 \left(\frac{Kd}{2}\right) =
 \frac{h_2'(z_+)}{W_h}\left[h_1(z_+) + S q h_2(z_+)\right]
\end{equation}
or Eq.~(\ref{eq:disp_law_sin})
\begin{equation}\label{eq:dispersion_law_smooth_modulation_sin}
\sin^2 \left(\frac{Kd}{2}\right) =
 -\frac{h_2(z_+)}{W_h}\left[h_1'(z_+) + S q h_2'(z_+)\right].
\end{equation}
These dispersion equations can be used to describe waves of all
polarizations by choosing appropriate for a given polarization
parameters $S$, $q$, and $h_{1,2}$. The band structure of a
passive photonic crystal can be obtained from these equations by
setting $\chi \equiv 0$. The band boundaries, $\Omega_\pm$, in
this case are determined by equations
\begin{equation}\label{eq:PC_gap_smooth_modulation}
h_1(z_+,\Omega_-) = 0, \qquad h'_2(z_+,\Omega_+) = 0,
\end{equation}
and
\begin{equation}\label{eq:PC_gap_smooth_modulation1}
h'_1(z_+,\Omega_-) = 0, \qquad h_2(z_+,\Omega_+) = 0,
\end{equation}
where $\Omega_\mp$ in the argument denote that the functions
$h_{1,2}$ are solutions of corresponding homogenous equations
[Eqs.~(\ref{eq:Maxwell_smooth_modulation_s_pol}) or
(\ref{eq:E_x_equation_p_pol})  with $\chi \equiv 0$] for $\omega =
\Omega_\mp$ respectively. Thus the zeroes of even and the maxima
of odd solutions given by Eqs.~(\ref{eq:PC_gap_smooth_modulation})
and (\ref{eq:PC_gap_smooth_modulation1}) determine the boundaries
of the consecutive band gaps in the band structure of the system
under consideration (see Fig.~\ref{fig:gaps_oscillating_theorem}).

The analysis of the full dispersion equations, with the resonance
terms restored, repeats the analysis of the previous subsection,
where one needs to make substitutions $a_{PC} = g_2$ and $f_{PC} =
g_1$. In particular, the parameters $t_\pm$ of the equation for
the boundaries of the forbidden gap,
Eq.~(\ref{eq:Bragg_gap_edges_eqs_mismatch}), are defined in terms
of boundary values of $f_{1,2}$ as
\begin{equation}\label{eq:T_pm_expressions_smooth}
 t_- = \frac{1}{\sqrt{W_h}}\left|\frac{\partial
h_1(\Omega_-)}{\partial\omega}\right|, \quad
 t_+ =\frac{1}{\sqrt{W_h}}\left|\frac{\partial
h_2'(\Omega_+)}{\partial\omega}\right|.
\end{equation}

The general structure of the spectrum is similar to the one
discussed in the case of the piece-wise modulation of the
refractive index, and we can generalize the condition for the
Bragg resonance, when the exciton related allowed band collapses,
and the spectrum of our system in the vicinity of the exciton
frequency consists of a single wide band gap. We again require
that the singularity of the excitonic susceptibility at $\omega_0$
is canceled by the first term $h_2'(z_+)$ in
Eq.~(\ref{eq:dispersion_law_smooth_modulation}). This corresponds
to the condition $\omega_0=\Omega_+$, where $\Omega_+$ is the high
frequency boundary of the respective photonic gap. From the
dispersion equation (\ref{eq:PC_gap_smooth_modulation}) it follows
again that the Bragg condition can be casted in the form
$K_{PC}(\omega_0)d=\pi$. One can rewrite the last equation in yet
another form emphasizing the role of the phase in the formation of
Bragg structures. Let us consider the effective optical width of
the period of the structure, $\tilde\phi$ [see
Eq.~(\ref{eq:relation_af_effective})]. At the point where $h_2' =
0$ one finds that $a/\bar{a} = -1$ yielding $\tilde\phi(\Omega_+)
= \pi$. Thus, the phase form of the Bragg condition, which is
particularly convenient for practical calculations, becomes
$\tilde\phi(\omega_0) = \pi$.

We can also generalize the expression for the width of the band
gap in the case of the Bragg resonance. Performing the expansion
similar to Eqs.~(\ref{eq:apprs_a_b}) we can obtain the equation
for boundaries of the forbidden gap:
\begin{equation}\label{eq:zeros_smooth_modulation}
(\omega-\Omega_+)\left(\omega - \Omega_- -
\frac{\tilde{\Delta}_\Gamma^2/4}{\omega-\omega_0}\right)= 0,
\end{equation}
with $\tilde{\Delta}_\Gamma^2$ given by the first part of
Eq.~(\ref{eq:PCMQW_effective_Delta}). The analysis of
Eq.~(\ref{eq:zeros_smooth_modulation}) completely repeats that of
Eq.~(\ref{eq:Bragg_gap_edges_eqs_mismatch}). In particular, the
width of the forbidden gap is determined by the Pythagorean sum of
the photonic and excitonic contributions,
\begin{equation}\label{eq:gap_width_smooth_modulation}
\Delta^2 = \Delta^2_{PC} + \tilde{\Delta}^2_\Gamma,
\end{equation}
where $\Delta_{PC} = |\Omega_+ - \Omega_-|$, and expression for the
excitonic contribution, $\tilde{\Delta}_\Gamma$, can be presented as
\begin{equation}\label{eq:modification_smooth_MQW_gap}
  \left(\frac{\tilde{\Delta}_\Gamma}{{\Delta}_\Gamma}\right)^2 =
  \frac{\pi c^2q_s}{4\Omega_- \omega_0 u^{(1)}_s} \quad
  \text{and} \quad \left(\frac{\tilde{\Delta}_\Gamma}{{\Delta}_\Gamma}\right)^2 =
  \frac{\pi c^2q_p p(z_+)}{4\Omega_- \omega_0 u^{(1)}_p}
\end{equation}
for $s$- and $p$-polarizations, respectively. Here we have
introduced
\begin{equation}\label{eq:s_p_even_energies}
  u^{(1)}_{s,p} = \frac 1 2 \int dz\, \epsilon(z) {\mathbf{E}_{s,p}^{(1)}}^2,
\end{equation}
where $\mathbf{E}_{s,p}^{(1)}$ is the $s$- and $p$-polarized
electric fields corresponding to the even mode of the photonic
crystal. This expression represents the energy associated with the
even mode of the photonic crystal concentrated in a single
elementary cell. The r.h.p.~of
Eqs.~(\ref{eq:modification_smooth_MQW_gap}) can be shown to be
proportional to $u^{(1)}_{qw}/u^{(1)}$, where $u^{(1)}_{qw}$ is
the energy of the even mode concentrated within the QW. This means
that the contribution of the exciton-light interaction into band
structure of the resonant photonic crystal depends on the
distribution of the energy of the electric field inside the
elementary cell at the frequency corresponding to $\Omega_-$. The
greater amount of energy stored within the well the larger the
role of the excitonic effects is.

\subsubsection{Off-Bragg structures}

Our analysis so far was limited to consideration of the band
boundaries and the properties of the band gaps in the resonant
photonic crystals. To complete the consideration of the normal
modes in these structures we need to discuss solutions of
Eqs.~(\ref{eq:PC_gap_smooth_modulation})  and
(\ref{eq:PC_gap_smooth_modulation}) in the allowed bands of the
spectrum. These solutions determine dispersion laws $\omega(K)$ of
the photonic crystal polaritons. In the case of the Bragg
structures and for frequencies in the vicinity of the band edges,
the dispersion laws can be presented in the following form:
\begin{equation}\label{eq:dispersion_curves_Bragg}
  \omega_\pm(K) = \Omega_c \pm
  \frac{1}{2}\sqrt{\Delta^2 +\zeta^2(K - K_B)^2},
\end{equation}
where $K_B = \pi/d$ is the Bloch wave number corresponding to the
boundary of the first Brillouin zone and $\zeta = d/t_+t_-$. These
two branches with positive and negative masses, $m_\pm = \pm
2\hbar^2\Delta/\zeta^2$, lie in the bands that are above and below
the forbidden gap, respectively. In addition to these branches
there is one more dispersionless branch at the exciton frequency
$\omega_B(K) = \omega_{0}$.

For the systems detuned from the exact Bragg resonance, the
positions of the band boundaries and properties of the band gaps
can be determined by repeating the arguments from the previous
sub-section; we do not reproduce it here. We however, complement
that analysis by considering modifications of the dispersion laws
of the polariton branches for the off-Bragg structures. If the
detuning is not too large, $|\delta|=|\omega_{0} -\Omega_+|
\lesssim \tilde{\Delta}_\Gamma$, the dispersion laws of the
original (present in the Bragg structure) upper and lower
polariton branches near the band edges can be written in the
following form
\begin{equation}\label{eq:off-Bragg_disp_law}
    \omega_\pm(K) = \omega'_\pm+\hbar^2 (K-K_b)^2/2m_\pm,
\end{equation}
where $\omega'_\pm$ are given by Eq.~(\ref{eq:sec_II_roots_gap})
with parameters defined by generalized equations
(\ref{eq:PC_gap_smooth_modulation}),
(\ref{eq:zeros_smooth_modulation}),  and
(\ref{eq:modification_smooth_MQW_gap}). The renormalized mass
parameters $m_\pm$ are defined as
\begin{equation}\label{eq:masses_positive_negative_off_Bragg}
  m_\pm(\delta) = m_\pm\left(1 - \frac{\Delta_{PC}\delta}{\Delta^2}
  \pm \frac{2\delta}{\Delta \mp \Delta_{PC}}\right).
\end{equation}
Comparing this result with Eq.~(\ref{eq:dispersion_curves_Bragg})
one can notice that detuning from the Bragg resonance results in
two main modifications of the dispersion laws: first, the position
of the band boundaries have changed and, second, the magnitudes of
the masses of the upper and the lower polariton branches are no
longer equal to each other.

The most dramatic changes occur, however, with the third, originally
dispersionless branch. In off-Bragg structures this branch acquires
dispersion, which, of course, agrees with the opening of the allowed
band in the vicinity of the exciton frequency. The dispersion law
characterizing this band is given by
\begin{equation}\label{eq:dispersion_curve_third}
 \omega_B(K) = \Omega_+ +
 \frac{\zeta^2\delta}{\tilde{\Delta}^2_\Gamma}(K - K_B)^2.
\end{equation}
This branch corresponds to excitations dubbed ``Braggaritons" in
Ref.~\onlinecite{Raikh_braggaritons}. The mass of this mode, given
by $m_B = \tilde{\Delta}_\Gamma^2/2\zeta\delta$, is very sensitive
to the amount of detuning from the Bragg condition and can be
effectively controlled, for instance, by the electric field via the
quantum confined Stark effect. This property of the "Braggariton"
branch invited proposal to use it for slowing, stopping and storing
light in BMQW
structures.\cite{Slow_lightBinderJOSA2005,{Slow_lightBinderOL2005}}

\subsubsection{Homogenous broadening}
We finish our consideration of dispersion properties of RPCs with
an arbitrary periodic distribution of the refractive index by
discussing effects due to the exciton homogeneous broadening. In
the presence of dissipation the concept of band gaps becomes
ill-defined because the imaginary part of the Bloch wave-number,
generally speaking, is not zero at all frequencies. Nevertheless,
the properties of this imaginary part are of great interest, since
they determine the spectral and transport properties of the
structures under consideration in the vicinity of what would have
been band boundaries in the absence of dissipation. In order to
access these properties we have to
consider solutions of Eq.~%
(\ref{eq:dispersion_law_smooth_modulation}). Taking into account
that in the spectral region of interest, the real part of $K$ is
close to $\pi/d$, we look for solutions to this equation in the
form
\begin{equation}\label{eq:K_representation}
  K d = \pi + i \lambda.
\end{equation}
\begin{figure}
\includegraphics[width=3.3in]{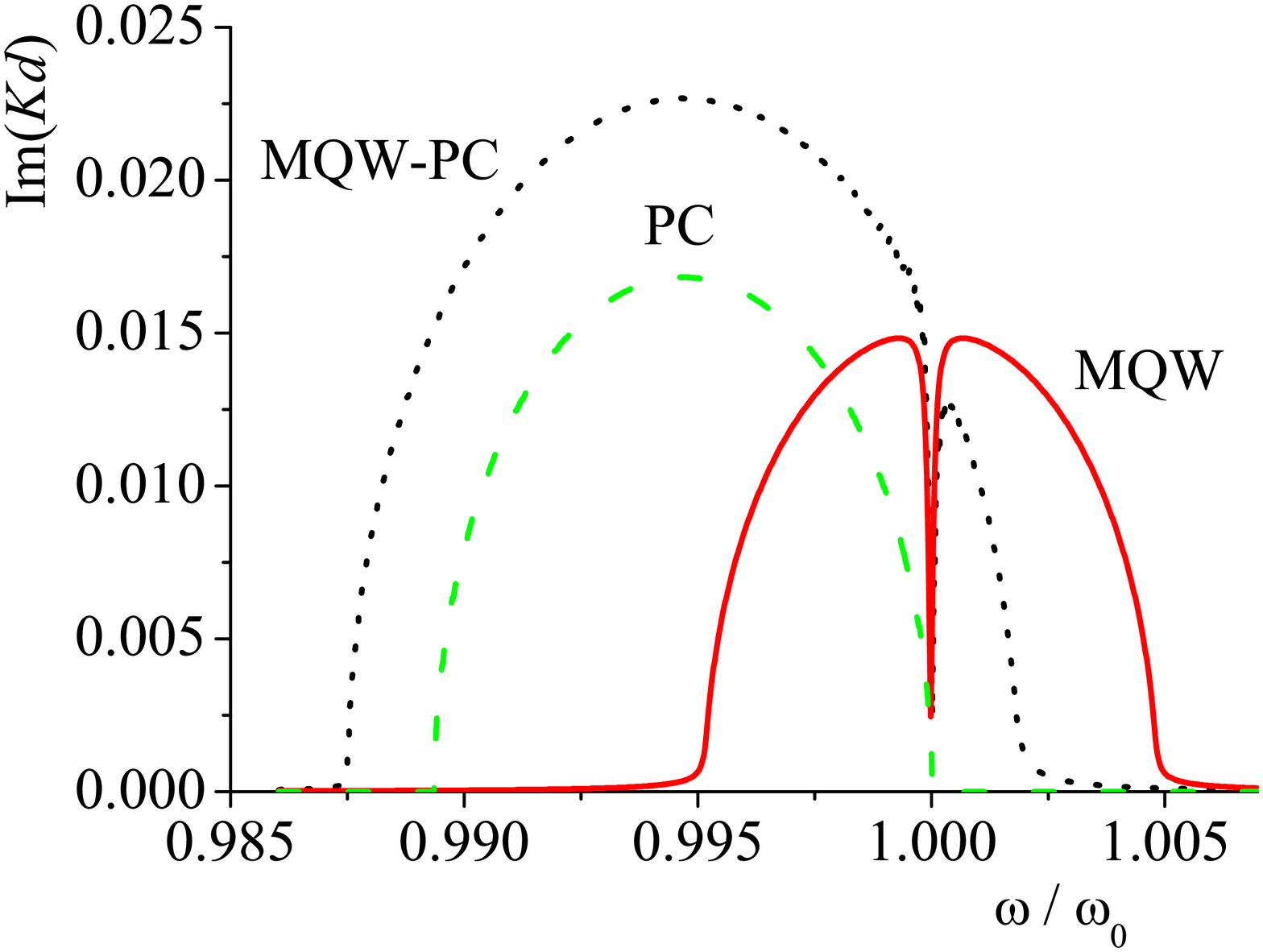}\\
\caption{Dependence of $\mathrm{Im}(Kd)$ on frequency for
different structures in the vicinity of the first forbidden gap
when the exciton homogeneous broadening is taken into account.
Dashed line represents the passive structure from
Fig.~\ref{fig:gaps_oscillating_theorem}. Solid line shows a Bragg
MQW  structure with a homogeneous dielectric function. Dotted line
represents  the structure with combination of QWs and the smooth
modulation of  the dielectric function. The characteristic feature
is a divergence of the penetration length $(\mathrm{Im}\,K)^{-1}$
at the exciton frequency.}\label{fig:comparison_structures_gaps}
\end{figure}
The unknown parameter $\lambda$ here has a simple physical
meaning: the inverse of its real part determines a characteristic
length on which the amplitude of the mode decays. The same length
determines the penetration depth of incident radiation, therefore
it is often called a penetration length. Assuming that $\lambda$
is small we can find an approximate expression for it in the form:
\begin{equation}\label{eq:solution_k}
 \lambda^2 = 4t_+t_-(\Omega_+ - \omega)\left(\omega - \Omega_-
 - \frac{\tilde{\Delta}_\Gamma^2/4}{\omega - \omega_0 - i\gamma}
 \right).
\end{equation}
An important feature described by this expression is that
$\lambda$ vanishes at the frequency corresponding to the right
edge of the photonic band gap, signifying the divergence of the
respective penetration length. The similar effect takes place in
MQW\cite{DLPRBSpectrum} structures without the refractive index
contrast. The important difference is that in the latter case the
divergence occurs at the exciton frequency that lies at the center
of the forbidden gap rather than at the band's boundary.
Fig.~\ref{fig:comparison_structures_gaps} shows the comparison of
the solutions of the dispersion equation for different structures
with the exciton homogenous broadening taken into account.

\subsection{MQW structures with complex elementary cells}
\label{sec:MQW_complex}

In this section we illustrate the application of the developed
technique to structures with complex elementary cells, which were
first studied in Ref.~\onlinecite{JointComplex}. In that paper the
mismatch between refractive indexes of different elements of the
structure had to be neglected because the combination of the
complex elementary cell and the spatial modulation of the
refractive index turned out to be an insurmountable obstacle for
the standard transfer-matrix approach. In this section we show
that the approach developed in the present paper allows us to
overcome the technical difficulties associated with the
consideration of a complex structure, and to generalize the
results of Ref.~\onlinecite{JointComplex} for the case of
structures with modulated refractive index.

\begin{figure}[tbp]
 \includegraphics[width=3in]{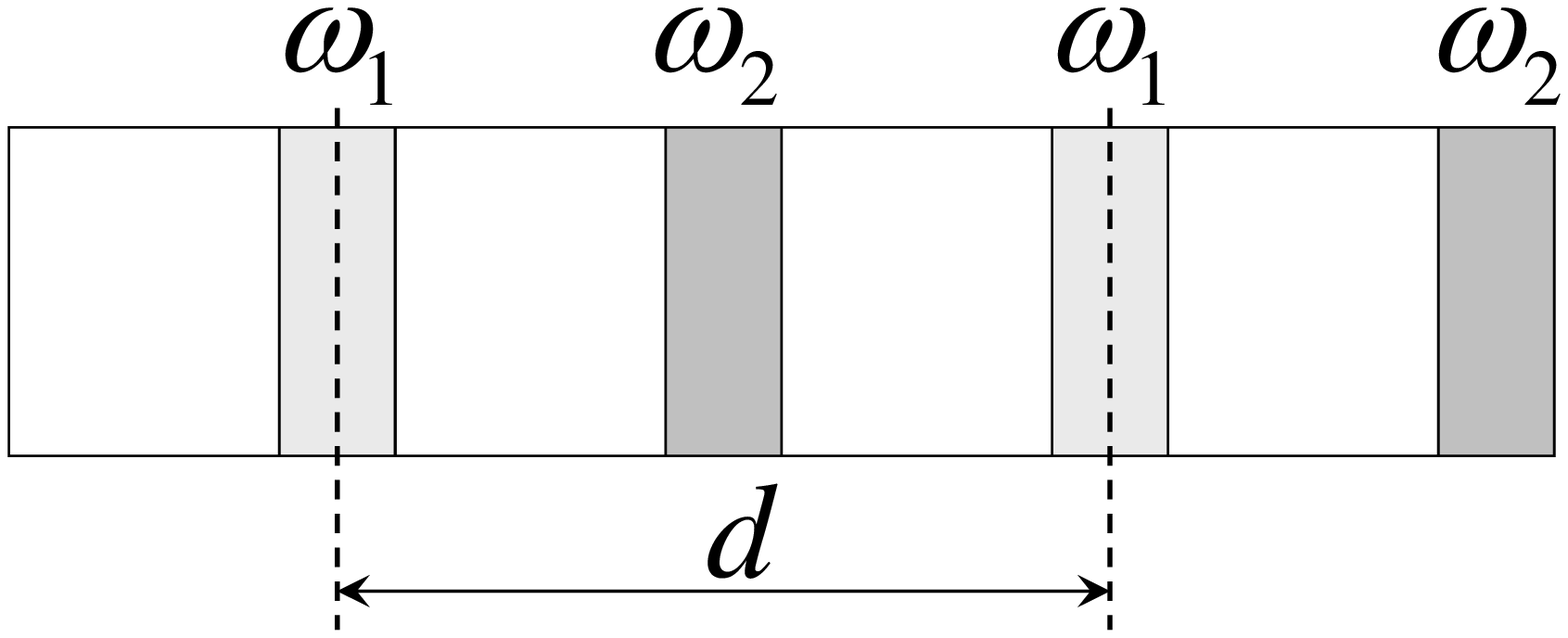}\\
 \caption{The periodic structure
with two QWs (dark rectangles) in the elementary cell. Dash lines
show the boundaries of the elementary cell having the mirror
symmetry. The QW with the exciton frequency $\omega_2$ is assumed
to have the index of refraction different from other elements of
the structure. }\label{fig:complex_cell}
\end{figure}

We will focus here on one particular example, when the elementary
cell consists of two QWs with different exciton frequencies,
$\omega_1$ and $\omega_2$, located half a period apart from each
other (see Fig.~\ref{fig:complex_cell}). It was found in
Ref.~\onlinecite{JointComplex} that despite the presence of two
different exciton frequencies, it is still possible to generalize
the notion of the Bragg resonance for such systems and design
structures whose spectrum would consist of only two polariton
branches separated by a wide band gap. However, unlike the case of
structures with a simple elementary cell, the formation of such a
wide band gap requires not only the period of the structure to
have a certain value, but also imposes a condition on the spectral
separation between the excitonic frequencies of the wells
constituting the elementary cell. When both generalized Bragg
conditions are fulfilled for such a structure, the width of the
polariton band gap, $\Delta_{CS}$, becomes larger than that in the
case of structures with a simple elementary cell:
\begin{equation}\label{eq:gap_complex_no_contrast}
\Delta_{CS} = \sqrt{2}\Delta_\Gamma.
\end{equation}
This broadening of the gap by almost $40\%$ reflects a possibility
to effectively strengthen the exciton-photon interaction by
increasing the density of QWs in the structure.

In this section we consider how the mismatch of the refractive
indices affects the spectral properties of such structures. We
simplify our consideration assuming that all elements of the
structure except the wells with the exciton frequency $\omega_2$
have the same refractive index. Formally, the dispersion equation
describing modes propagating in such a structure in the normal
direction has been obtained in Ref.~\onlinecite{JointComplex} by
more conventional methods, but that equation turned out to be too
cumbersome to allow for any non-numerical analysis. Using
$(a,f)$-representation of the transfer matrices, we show here that
the dispersion equation can be rewritten in a much more
transparent form with a factorized right hand side.

To apply the technique developed in the present paper it is
necessary to choose the elementary cell with the explicit mirror
symmetry. It can be done as shown in Fig.~\ref{fig:complex_cell}.
The problem of determining the transfer matrices through the right
and left halves of the QW can be resolved in the following way. We
note that the QW transfer matrix $T_w$ determined by
Eq.~(\ref{eq:transfer_matrix_well}) can be written in a factorized
form
\begin{equation}\label{eq:transfer_matrix_well_factorized}
T_w = T_b(\phi_w/2)\tilde{T}_w T_b(\phi_w/2),
\end{equation}
where $\tilde{T}_w$ is derived from the expression for $T_w$ by
setting $\phi_w = 0$. After this factorization the transfer matrix
through the elementary cell takes the form similar to
Eq.~(\ref{eq:simple_sandwich_transfer})
\begin{equation}\label{eq:contrast_transfer_period_mirror_symmetry}
 T = \sqrt{\tilde{T}^{(1)}_w}\, T_b(\phi_b + \phi_w/2)\, T^{-1}_\rho T^{(2)}_w\, T_\rho  T_b(\phi_b + \phi_w/2)\,
\sqrt{\tilde{T}^{(1)}_w}.
\end{equation}
The difference between the indices of refraction is again taken
into account by introducing the special transfer matrix $T_\rho$,
which describes propagation of the wave across the interface and
contains respective Fresnel coefficients. The transfer matrices
through different QWs, $T_w^{(1,2)}$, are obtained from
Eq.~(\ref{eq:transfer_matrix_well}) by substituting different
excitonic susceptibilities
\begin{equation}\label{eq:wells_susceptibilities}
S_{1,2} = \frac{\Gamma_0}{\omega - \omega_{1,2}+i\gamma}.
\end{equation}
The matrix square root $\sqrt{\tilde{T}^{(1)}_w}$ can be found to be
equal to
\begin{equation}\label{eq:well_transfer_decomposition_center}
\sqrt{\tilde{T}^{(1)}_w} =
\begin{pmatrix}
 1 - i S_1/2 & -iS_1/2 \\
 iS_1/2 & 1 + i S_1/2
\end{pmatrix}.
\end{equation}

Now, we can apply
Eqs.~(\ref{eq:simple_sandwich_transfer_elements}),
(\ref{eq:sandwich_barriers}) and (\ref{eq:sandwich_mismatch}) to
derive $(a,f)$-representation for $T$. The dispersion equation
following from this representation has a relatively simple form:
\begin{equation}\label{eq:complex_contrast_dispersion_law}
\cos^2\left(\frac{Kd}{2}\right) = \frac{1}{1-\rho^2}
\left[\mathrm{Re}(a) + S_1\, \mathrm{Im}(a) \right]
\left[\mathrm{Re}(b) + S_2\, \mathrm{Im}(b) \right],
\end{equation}
where $a = \exp(i\phi_+) - \rho\exp(i\phi_-)$, $b = \exp(i\phi_+) +
\rho\exp(i\phi_-)$ [compare with
Eqs.~(\ref{eq:af_parameters_passive_multilayer})], and $\phi_\pm =
\omega (2 d_b n + d_w n \pm d_w n_2)/2c$. 

As usual, the positions of band boundaries are determined by
frequencies at which the r.h.s. of
Eq.~(\ref{eq:complex_contrast_dispersion_law}) changes its sign,
which can occur either by passing this expression through zero or
through infinity. The other set of boundaries, which we do not
consider here can be obtained as before by converting
Eq.~(\ref{eq:complex_contrast_dispersion_law}) to the form with
$\sin^2\left(\frac{Kd}{2}\right)$ on the left. We can show that in
the immediate proximity of the exciton frequencies, there are six
band boundaries, which define three band gaps separated by two
allowed bands. We again will be interested in finding conditions,
under which these allowed bands collapse resulting in formation of
one continuous band gap. As we already know from the previously
considered examples, the collapse of the allowed bands associated
with the singularities of the exciton susceptibility, $S$, takes
place when the exciton frequency coincides with one of the band
boundaries of a certain effective structure. In the case of a
simple elementary cell this effective structure was a respective
passive photonic crystal, however, in the case under
consideration, the effective structure is comprised of one of the
sublattices of our system. More precisely, in order to cancel the
exciton singularity at, for example, $\omega_1$, we have to
require that $\omega_1$ coincides with the zero of the expression
inside the second parenthesis on the r.h.s. of
Eq.~(\ref{eq:complex_contrast_dispersion_law}). This zero,
obviously gives the position of some band boundaries of a
structure made up of the sub-lattice 2. Similarly, in order to
cancel the singularity at $\omega_2$ we need to make it equal to
one of the band boundaries originating from the sub-lattice 1,
particulary that which turns to zero the expression in the first
parenthesis of the r.h.s of
Eq.~(\ref{eq:complex_contrast_dispersion_law}). Let us assume for
concreteness that $\omega_2 > \omega_1$, then more specifically we
have to require that $\omega_1$ is equal to the low-frequency
boundary of the band gap associated with polariton branches in
sub-lattice 2, and $\omega_2$ should coincide with the
high-frequency boundary of the band gap originating from polariton
branches in sublattice 1. Let us denote frequencies of these two
band boundaries as $\omega^\pm_{1,2}$, where the lower index shows
from which sublattice the respective boundary originates, and
$\pm$ denotes high- and low-frequency boundaries, respectively.
The frequencies $\omega^\pm_{1,2}$ can be found to be equal to
\begin{equation}\label{eq:complex_characteristic_frequencies}
\begin{split}
 \omega_1^\pm = \frac 1 2 (\omega_1 + \Omega_-) \pm \frac 1 2
 \sqrt{(\omega_1 - \Omega_-)^2 + \Delta_\Gamma^2}, \\
 \omega_2^\pm = \frac 1 2 (\omega_2 + \Omega_+) \pm \frac 1 2
 \sqrt{(\omega_2 - \Omega_+)^2 + \Delta_\Gamma^2},
\end{split}
\end{equation}
where $\Omega_\mp$ are determined by
Eqs.~(\ref{eq:gap_edges_PC_explicitely}), and for simplicity, we
assumed the same value of the effective exciton-light interaction
for both sublattices, i.e.~we neglected a possible difference
between exciton radiative decay rates in different wells. We also
neglected the renormalization of $\Gamma_0$ due to the spatial
modulation of the dielectric function [see
Eqs.~(\ref{eq:PCMQW_effective_Delta}) and
(\ref{eq:modification_smooth_MQW_gap})]. The generalization of
these expressions is straightforward but does not lead to anything
new while making the resulting expressions much more cumbersome.

Let us, first, consider the case when the index of refraction of the
wells of the second type is higher, i.e. $\rho >0$. The conditions
of absence of transparency windows inside the gap (the Bragg
resonance conditions) are in this case formulated as $\omega_1^+ =
\omega_2$ and $\omega_2^- = \omega_1$. Resolving these equations
with respect to the exciton frequencies one obtains
\begin{equation}\label{eq:complex_Bragg_condition}
  \frac{\omega_1+\omega_2}{2} = \Omega_c, \qquad
  \omega_1-\omega_2 = \frac 1 2 \left(\sqrt{2 \Delta_\Gamma^2 +
  \Delta_{PC}^2}- \Delta_{PC}\right),
\end{equation}
which generalize the results of Ref.~\onlinecite{JointComplex} to
the case of structures with the refractive index mismatch.
Substituting these results into the expression for the width of
the band gap, which is equal to $\Delta = \omega_2^+ -
\omega_1^-$, we obtain a familiar result
\begin{equation}\label{eq:complex_gap_width}
 \Delta = \sqrt{\Delta_{CS}^2 + \Delta_{PC}^2}.
\end{equation}
The case of $\rho<0$ is described by the same formulas with
substitution $\Delta_{PC} \rightarrow -\Delta_{PC}$, which changes
the Bragg condition [the second equation of
Eqs.~(\ref{eq:complex_Bragg_condition})] but leaves the width of
the gap, Eq.~(\ref{eq:complex_gap_width}), the same.

Thus, taking into account the mismatch of the indices of refraction
in the system with a complex elementary cell leads to modification
of the Bragg condition in comparison with what one has in the case
without the mismatch. This modification has a different character
than in structures with a simple cell. The mean value of the exciton
frequencies stays at the center of the forbidden gap and only the
difference between the frequencies must be corrected to take into
account the mismatch. The resultant gap is, similar to the case of
simple cell, again the Pythagorean sum of the photonic and the
excitonic contributions.

\section{Conclusion}

We considered the band structure associated with normal
excitations of one-dimensional resonant photonic crystals using as
a particular example the exciton polaritons multiple quantum well
structures. Properties of normal modes of such structures are
determined by the interplay of two mechanisms: the interaction of
light with internal resonances of constituting materials, such as
excitons in MQW case, and multiple reflection of light due to
periodic spatial modulation of the index of refraction of the
structure. The necessity to take into account both mechanisms of
the light-matter interaction in these structures on equal footing
makes the analytical description of their properties more
complicated than in the case of regular passive photonic crystals.
In this paper we developed a powerful method of analyzing band
structure and dispersive properties of such structures, which
allowed us to obtain analytical solutions for problems that could
not be solved by other approaches.

This method is based on two key features. First, we construct
transfer matrices using solutions of the appropriate Cauchy
(rather than boundary value) problems as a basis. This makes
calculations of the transfer matrices more convenient since it
allows for avoiding difficult questions about eigenmodes of the
system under consideration. Second, we use a special
representation of the transfer matrix valid for structures with an
elementary cell possessing a mirror symmetry. This representation
automatically yields the dispersion equation of normal modes in a
factorized form, which drastically simplifies the analysis of the
spectrum. In particular, using this representation we were able to
analyze dispersion properties of certain type of resonant photonic
crystals with complex elementary cells. We also believe that
besides situations explicitly considered in the paper, the
suggested method allows for effective treatment of other problems
such as exciton luminescence from MQW photonic crystals, or for
analysis  of more complicated systems, e.g. quantum
graphs.\cite{Kuchment:2002,Kuchment:2004} An important benefit of
the developed approach consists also in the fact that polarization
and angular dependencies of the elements of transfer matrices
appear in this approach only via respective Fresnel coefficients.
This enabled us to obtain band structures of modes of both $s$-
and $p$-polarizations in MQW based photonic crystals from
essentially the same dispersion equation and easily analyze their
angular dependencies.

We illustrated the developed method by applying it to a number of
problems, some of which were considered previously in the
framework of other approaches, while others were analyzed in this
paper for the first time. In particular, we considered polariton
spectrum in several systems: MQW structures with the mismatch of
the indices of refraction of the barriers and wells materials, MQW
with an arbitrary periodic modulation of the dielectric function,
and MQW with a complex elementary cell consisting of two
equidistant QWs characterized by different exciton frequencies. We
showed that in all these cases one can construct a structure, in
which multiple photon and polariton bands can be collapsed in just
two polariton branches separated by an anomalously broad stop
band. Such a band structure is characteristic of so called Bragg
structures, which in the case of MQWs were studied in a number of
previous publications. We showed here that in the case of RPCs
with a simple elementary cell the condition for formation of such
a spectrum can be formulated as a requirement for the exciton
frequency to coincide with one of the edges of the respective
passive photonic crystal. Formulated in this form this condition
demonstrates a direct connection with earlier formulations of the
Bragg resonance in MQW structures without the refractive index
contrast. In the case of crystals with complex elementary cells,
we showed that our method allows for taking into account the
mismatch of the refractive indices between different elements of
the structure and found the generalized Bragg conditions for this
system as well. In the last case, this condition consists of two
equations: one relating the period of the structure to the exciton
frequencies of the constituent elements and the other specifying a
relation between those frequencies.

One of the surprising results is that for all these structures the
width of the polariton band gap in the case of the Bragg resonance
can be presented in the form of a Pythagorean sum of the photonic
and the excitonic contributions, $\Delta^2 = \Delta_{PC}^2 +
\tilde{\Delta}_\Gamma^2$. The excitonic contribution in the
presence of the refractive index modulation, however, is modified
compared to the case of simple optical lattices. It is interesting
that this modification can be presented in the form of a ratio of
the electric field energy concentrated inside QW to the the total
energy stored in the entire elementary cell of the structure.

\acknowledgments

The work has been supported by AFOSR grant F49620-02-1-0305.

\appendix
\section{Transfer matrix formalism}

The transfer matrix technique has been reviewed in a number of
publications. Here, however, we use this technique in a somewhat
unusual form. Therefore, we find it useful to recall some
important results. While the general idea of the transfer matrix
establishing a linear connection between values of a function
representing a solution of a linear differential equation at two
different values of the argument is inherent to all formulations,
the explicit form of the transfer matrix and its properties depend
upon a choice of the representation of the matrix. As a basic
representation one often chooses a $2\times 2$ matrix operating on
a two-dimensional vector, whose components are the value of a
function, which represents a solution at a given point, and its
derivative. The formal reason for this choice is the fact that a
linear equation of a second order can always be naturally
reformulated as a  system of two first order equations for the
function and its derivative:
\begin{equation}\label{eq:transfer_function_derivative_basis}
\begin{pmatrix}  E(z+d) \\  E'(z+d) \end{pmatrix} =
T_\psi(z+d, z)
\begin{pmatrix}  E(z) \\  E'(z)
\end{pmatrix}.
\end{equation}

Another approach to constructing transfer matrices is more
convenient for problems involving scattering of  waves. This
problem is most naturally described by specifying an amplitude of
an incident wave and relating to it amplitudes of transmitted and
reflected waves. In this approach, the field is sought in the form
$e^{i\kappa z} + r e^{-i\kappa z}$ at the one side of the system,
and in the form $te^{i\kappa z}$ at the the other side. More
generally, the field is represented in the form
\begin{equation}\label{eq:wave_basis}
\begin{split}
E(z)= a_+ e^{i\kappa z} + a_- e^{-i\kappa z}, \\
E(z+d)= a'_+ e^{i\kappa z} + a'_- e^{-i\kappa z}
\end{split}
\end{equation}
and the transfer matrix relates the amplitudes at the left and the
right boundaries
\begin{equation}\label{eq:transfer_matrices_waves_basis}
\begin{pmatrix}
a'_+ \\ a'_-
\end{pmatrix}= T \begin{pmatrix} a_+ \\ a_- \end{pmatrix}.
\end{equation}
As follows from Eqs.~(\ref{eq:wave_basis}), the transfer matrices
written in these two bases  are related by a similarity
transformation, $T = W_e^{-1}(z)T_\psi(z+d,z)W_e(z)$, where
\begin{equation}\label{eq:relation_matrices_field_waves}
 W_e(z) =
\begin{pmatrix}
 e^{i\kappa z} & e^{-i\kappa z} \\
 i\kappa e^{i\kappa z} & -i\kappa e^{-i\kappa z}
\end{pmatrix}.
\end{equation}

Another important way to introduce the transfer matrix (see e.g.~%
Ref.~\onlinecite{PEREZ-ALVAREZ:2001}) is based on a possibility of
a representation of the solution of the Maxwell equation as a sum
of two linearly independent functions, $f_{1,2}(z)$, with
modulated amplitudes
\begin{equation}\label{eq:general_basis_definition}
\begin{split}
E(z) = c_1(z) h_1(z) + c_2(z) h_2(z), \\
E'(z) = c_1(z) h'_1(z) + c_2(z) h'_2(z).
\end{split}
\end{equation}
Now, the transfer matrix gives the relation between the amplitudes
at different points, usually a period of a structure apart,
\begin{equation}\label{eq:transfer_general_basis}
\begin{pmatrix}
c_1(z+d) \\  c_2(z+d)
\end{pmatrix} = T_h(d) \begin{pmatrix}
c_1(z) \\  c_2(z) \end{pmatrix}.
\end{equation}
>From Eq.~(\ref{eq:general_basis_definition}) it follows that
\begin{equation}\label{eq:relation_field_functions}
T_h(d) = W_h^{-1}(z+d)T_\psi(z+d,z) W_h(z),
\end{equation}
where $W_h(z)$ is the Wronsky matrix
\begin{equation}\label{eq:Wronsky_matrix_definition}
W_h(z) = \begin{pmatrix}    h_{1}(z) & h_{2}(z) \\
h'_{1}(z) & h'_{2}(z)  \end{pmatrix}.
\end{equation}
Eq.~(\ref{eq:Wronsky_matrix_definition}) allows one to derive a
relation between transfer matrices obtained for a different choice
of the basis functions $h_{1,2}$.

The relation between the transfer matrices written in the bases of
plane waves and a pair of linearly independent functions can be
written as
\begin{equation}\label{eq:relation_transfer_matrices}
T = M(d/2)T_h(d)M^{-1}(-d/2),
\end{equation}
where $M(z) =W_e^{-1}(0)W_h(z)$. Using
Eqs.~(\ref{eq:relation_matrices_field_waves}),
(\ref{eq:Wronsky_matrix_definition}), and
(\ref{eq:relation_transfer_matrices}) one obtaines
\begin{equation}\label{eq:transition_to_pseudo-waves}
M(z) = \frac{1}{2}
\begin{pmatrix}
 h_{1}(z)+\dfrac{h'_1(z)}{i\kappa} &
h_{2}(z)+\dfrac{h'_2(z)}{i\kappa} \\
 h_{1}(z)-\dfrac{h'_1(z)}{i\kappa} &
h_{2}(z)-\dfrac{h'_2(z)}{i\kappa}
\end{pmatrix}.
\end{equation}

All types of transfer matrices have their advantages and
disadvantages. For instance, a basis of a pair of linearly
independent functions can be most naturally related to solutions
of the Cauchy problem for the respective differential equation.
Such a basis is easy to find but it is inconvenient for the
solution of scattering problems. The fact that, generally, it is
not related to $T_\psi$ by a similarity transformation [see
Eq.~(\ref{eq:relation_field_functions})] makes it also difficult
to consider the spectral problems using this basis. Indeed, the
knowledge of the transfer matrix itself is not sufficient for
solving this problem, and one has to separately consider a
boundary value problem for amplitudes $c_{1,2}$. On the contrary,
the basis of plane waves is the most suitable for solving this
particular problem since the transfer matrix itself contains all
needed information. In this basis, the difficulty of having to
solve the boundary value problem is moved to the process of
finding the transfer matrix itself. Therefore, the conversion rule
from one basis to another presented in
Eq.~(\ref{eq:relation_transfer_matrices}) is very useful for
practical calculations as it is demonstrated in our paper.


\end{document}